\documentclass[final,11pt,3p,doublespacing]{elsarticle}
\journal{.}
\usepackage{times}
\emergencystretch=\maxdimen
\usepackage{sectsty}
\sectionfont{\color{blue}\large\MakeUppercase}
\subsectionfont{\color{blue}}
\biboptions{sort&compress}
\usepackage[colorlinks]{hyperref}
\AtBeginDocument{%
	\hypersetup{
		plainpages = false, 
		bookmarksopen = true,
		bookmarksnumbered = true,
		breaklinks = true,
		linktocpage,
		colorlinks = true,
		linkcolor = purple,
		urlcolor  = black,
		citecolor = Magenta,
}}
\usepackage{multicol}
\usepackage{multirow}
\usepackage{imakeidx}
\makeindex
\usepackage{nomencl}
\makenomenclature
\usepackage[xindy]{glossaries}
\makeglossaries
\usepackage{graphics}
\usepackage{graphicx}
\usepackage{epsf,psfrag}
\usepackage{epstopdf}
\usepackage[para]{threeparttable}
\usepackage{subfigure}
\usepackage{epsfig}
\usepackage{pdflscape}
\usepackage{rotating}
\usepackage{lscape}
\usepackage{float}
\usepackage{stfloats}
\usepackage{textgreek}
\usepackage{longtable}
\usepackage{lscape}
\usepackage{capt-of}
\usepackage{tablefootnote}
\usepackage{footnote}
\usepackage{scrextend}
\usepackage{caption}
\usepackage[autopunct=true]{csquotes}
\usepackage{rotating}
\usepackage[normalem]{ulem}
\usepackage{resizegather}
\usepackage{siunitx}
\usepackage{cases}
\usepackage[nodisplayskipstretch]{setspace}
\usepackage{color,soul}
\usepackage[usenames,dvipsnames]{xcolor}
\usepackage{amsfonts,amsmath,amssymb,gensymb,stmaryrd}
\usepackage{latexsym,array}
\usepackage{textcomp}

\newcommand{\RomanNumeralCaps}[1]
\linenumbers
\makesavenoteenv{tabular}
\makesavenoteenv{table}
\usepackage[left,displaymath, mathlines]{lineno}
\newcommand{\overbar}[1]{\mkern 1.5mu\overline{\mkern-1.5mu#1\mkern-1.5mu}\mkern 1.5mu}
\DeclareMathAlphabet{\mathpzc}{OT1}{pzc}{m}{it}
\def\fig{Fig.~}
\def\figs{Figs.~}
\def\eqn{Eq.~}
\def\eqns{Eqs.~}
\def\tab{Table~}

\newcommand{\myvec}[1]{\mathbf{#1}}     

\def\tsc#1{\csdef{#1}{\textsc{\lowercase{#1}}\xspace}}
\tsc{WGM}
\tsc{QE}
\tsc{EP}
\tsc{PMS}
\tsc{BEC}
\tsc{DE}
\usepackage{easyReview} 
\newcommand{\removeEq}[1]{\ifistoreview{\@\expandafter\removeColor{#1}\hspace{-0.6em}} \else {}\fi}
     
\graphicspath{{../}{Figures/}{Figures/Validation/}}
\DeclareGraphicsExtensions{.png, .PNG,.pdf,.PDF,.jpg,.JPG,.eps,.EPS}
\begin{document}
\setcounter{page}{1}
\begin{frontmatter} 
%
%
%
%
%
%
\title{CFD analysis of electroviscous effects in electrolyte liquid flow through heterogeneously charged uniform microfluidic device}
\author[labela]{Jitendra {Dhakar}}
\author[labela]{Ram Prakash {Bharti}\corref{coradd}}
\emailauthor{rpbharti@iitr.ac.in}{R.P. Bharti}
\address[labela]{Complex Fluid Dynamics and Microfluidics (CFDM) Lab, Department of Chemical Engineering, Indian Institute of Technology Roorkee, Roorkee - 247667, Uttarakhand, India}
%
%
\cortext[coradd]{\textit{Corresponding author. }}
%
\begin{abstract}
\fontsize{11}{14pt}\selectfont
Charge-heterogeneity (i.e., surface charge variation in the axial direction of device) introduces non-uniformity in flow characteristics in the microfluidic device. Thus, it can be used for controlling the practical microfluidic applications, such as mixing, mass, and heat transfer processes. This study has numerically investigated the charge-heterogeneity effects in the electroviscous (EV) flow of symmetric ($1$:$1$) electrolyte liquid through a uniform slit microfluidic device. The Poisson's, Nernst-Planck (N-P), {and} Navier-Stokes (N-S) equations are numerically solved using the finite element method (FEM) to obtain the flow fields, such as total electrical potential ($U$), excess charge ($n^\ast$), induced electric field strength ($E_\text{x}$), and pressure ($P$) fields for following ranges of governing parameters: inverse Debye length ($2\le K\le 20$), surface charge density ($4\le  {S_\text{1}}\le 16$), and surface charge-heterogeneity ratio ($0\le  {S_\text{rh}}\le 2$). Results have shown that the total potential ($|\Delta U|$) and pressure ($|\Delta P|$) drop maximally increase by 99.09\% (from 0.1413 to 0.2812) (at $K=20$, $ {S_\text{1}}=4$) and 12.77\% ({from} 5.4132 to 6.1045) (at $K=2$, $ {S_\text{1}}=8$), respectively with overall charge-heterogeneity ($0\le  {S_\text{rh}}\le 2$). Electroviscous correction factor ($Y$, i.e., ratio of effective to physical viscosity) maximally enhances by 12.77\% ({from} 1.2040 to 1.3577) (at $K=2$, $ {S_\text{1}}=8$), 40.98\% ({from} 1.0026 to 1.4135) (at $ {S_\text{1}}=16$, $ {S_\text{rh}}=1.50$), and 41.35\% ({from} 1 to 1.4135) (at $K=2$, $ {S_\text{rh}}=1.50$), with the variation of $ {S_\text{rh}}$ (from 0 to 2), $K$ (from 20 to 2), and $ {S_\text{1}}$ (from 0 to 16), respectively. Further, a simple pseudo-analytical model is developed to estimate the pressure drop in EV flow, accounting for the influence of charge-heterogeneity based on the Poiseuille flow in a uniform channel. This model predicts the pressure drop $\pm$2--4\% within the numerical results. The robustness and simplicity of this model enable the present numerical results for engineering and design aspects of microfluidic applications.
\end{abstract}
\end{frontmatter}
\section{Introduction}
\label{sec:intro}
Recent advances in micro-fabrication technology, micro-electro-mechanical systems (MEMS), and biochemical devices have enhanced their uses in several fields, such as chemical, medical, and biological \citep{li2008encyclopedia,bhushan2007springer,lin2011microfluidics,nguyen2013design,bruijns2016microfluidic,han2020review,li2021microfluidic,laucirica2024advances}. Microfluidic devices increase the heat and mass transfer rates of the processes used in practical applications. Traditional theories used for macro-scale flows are not valid for micro-scale flows due to reduced dimensions, {as the surface} forces such as surface tension, magnetic field, electrical charges, etc., remarkably affect the micro-scale flows \citep{hunter2013zeta,li2001electro}. Understanding the electrokinetic phenomena is essential at the micro-scale analysis to develop reliable and efficient microfluidic devices for practical microfluidic applications.

Electrokinetic phenomena arise when charged solid surfaces interact with electrolyte liquid  (refer \fig\ref{fig:1}). The charged surfaces affect ion distribution near the solid-liquid interface, forming an \textit{electrical double layer} (EDL)  \citep{atten1982electroviscous,li2001electro,hunter2013zeta,dhakar2022slip,dhakar2022electroviscous,dhakar2023cfd}. It consists of Stern (or compact) and diffuse layers, separated by the shear plane. The potential at the shear plane is known as \textit{zeta potential} ($\zeta$), and it decays in the diffuse layer away from the surface. The counter-ions in the diffuse layer of EDL are {advected} by applied pressure-driven flow (PDF) along the downstream end, resulting in the \textit{streaming current} ($I_\text{s}$). Subsequently, the accumulation of ions along the length of the device results in the \textit{streaming potential}. It drives counter-ions in the diffuse layer of EDL in the opposite of PDF, resulting in the current known as \textit{conduction current} ($I_\text{c}$), generating a flow in the opposite direction of the primary PDF. This results in a reduction in the net flow rate in the PDF direction. This effect is commonly called  \citep{atten1982electroviscous,li2001electro,hunter2013zeta,dhakar2022slip,dhakar2022electroviscous,dhakar2023cfd} the \textit{electroviscous effect} (EVE).  

{Prior to discussing the relevant literature, it is helpful to define the (i)  surface charge density ratio ($S_\text{r}$) as the ratio of surface charge densities ($\sigma_\text{i}$) of the opposing walls/sections, (ii) surface charge heterogeneity ratio ($S_\text{rh}$) as the ratio of surface charge densities ($\sigma_\text{i}$) of the different uniformly charged sections ($1\le k\le n$) of the wall, expressed as follows.
\begin{gather}
	S_\text{r} = \frac{\sigma_{k,w=\text{b}}}{\sigma_{k,w=\text{t}}}
	; \qquad\text{and}\qquad 
	S_\text{rh}= \frac{\sigma_{{k},w}}{\sigma_{k=l,w}} 
	\label{eq:srrh}
\end{gather}
where, $w =(\text{b}, \text{t})$ indicates the walls/surfaces ($\text{b}$ for bottom, and $\text{t}$ for top); $k$ refers to the sections of the wall/surface.
Furthermore, $S_\text{r}=1$ and $S_\text{r}\neq1$ refer to the symmetrically and asymmetrically charged; whereas,  $S_\text{rh}=1$ and $S_\text{rh}\neq1$ refer to the homogeneously and heterogeneously charged microfluidic device, respectively.}
%

Recent studies \citep{dhakar2022electroviscous,dhakar2023cfd,dhakar2024influence} have reviewed the literature about electroviscous flow in the symmetrically 
($S_\text{r}=1$) and homogeneously 
($S_\text{rh}=1$) charged microfluidic devices of uniform cross-sections, such as slit \citep{burgreen1964electrokinetic,mala1997flow,mala1997heat,chun2003electrokinetic,ren2004electroviscous,chen2004developing,joly2006liquid,wang2010flow,jamaati2010pressure,zhao2011competition,tan2014combined,jing2015electroviscous,matin2016electrokinetic,jing2017non,matin2017electroviscous,kim2018analysis,mo2019electroviscous,li2021combined,li2022electroviscous,liu2024electrokinetic}, cylinder \citep{rice1965electrokinetic,levine1975theory,bowen1995electroviscous,brutin2005modeling,bharti2009electroviscous,jing2016electroviscous}, rectangular \citep{yang1998modeling,li2001electro,ren2001electro}, and elliptical \citep{hsu2002electrokinetic}, as well as non-uniform cross-sections, such as contraction-expansion slit \citep{davidson2007electroviscous,Berry2011,dhakar2022slip,dhakar2022electroviscous,dhakar2024influence}, cylinder \citep{bharti2008steady,davidson2010electroviscous}, and rectangular \citep{davidson2008electroviscous}. 
Further, \citet{dhakar2023cfd,dhakar2024icchmt} have analyzed the electroviscous effects in the electrolyte liquid flow through an asymmetrically 
 ($S_\text{r}\neq1$) and homogeneously 
 ($S_\text{rh}=1$) charged contraction-expansion slit microfluidic device. 
These studies \citep{davidson2007electroviscous,davidson2008electroviscous,bharti2008steady,bharti2009electroviscous,davidson2010electroviscous,Berry2011,dhakar2022electroviscous,dhakar2022slip,dhakar2023cfd,dhakar2024influence} have concluded that the surface charge density ($4\le S\le 16$), inverse Debye length ($2\le K\le 20$), surface charge ratio ($0\le  {S_\text{r}}\le 2$), and slip length ($0\le B_0\le 0.20$) significantly affect the hydrodynamic fields, such as total electrical potential ($U$), induced electric field strength ($E_\text{x}$), excess charge ($n^\ast$), and pressure ($P$) fields in the homogeneously ($S_\text{rh}=1$) charged microfluidic devices. Simple pseudo-analytical models have also been developed \citep{davidson2007electroviscous,davidson2008electroviscous,bharti2008steady,bharti2009electroviscous,davidson2010electroviscous,Berry2011,dhakar2022electroviscous,dhakar2022slip,dhakar2023cfd}, based on the Poiseuille flow in a uniform channel, to calculate the pressure drop ($\Delta P$) and electroviscous correction factor ($Y$, i.e., ratio of effective to physical viscosity), which estimate the pressure drop within the acceptable level ($\pm 5\%$) with their numerical results \citep{davidson2007electroviscous,davidson2008electroviscous,bharti2008steady,bharti2009electroviscous,davidson2010electroviscous,Berry2011,dhakar2022electroviscous,dhakar2022slip,dhakar2023cfd,dhakar2024influence}. {Broadly, the existing studies \citep{davidson2007electroviscous,davidson2008electroviscous,bharti2008steady,bharti2009electroviscous,davidson2010electroviscous,Berry2011,dhakar2022electroviscous,dhakar2022slip,dhakar2023cfd,dhakar2024influence} have used the homogeneously charged ($S_\text{rh}=1$) microfluidic devices, whereas the present study is based on the heterogeneously charged ($S_\text{rh}\neq 1$) microfluidic devices.}
%

Surface heterogeneity ($S_\text{rh}$) is an essential characteristic of the microfluidic device, which can arise due to surface treatment defects \citep{ajdari1996generation}, chemical species absorption by surface \citep{ghosal2003effect}, and controlling the surface charge distribution \citep{jain2013optimal,bhattacharyya2019enhanced}. Surface charge heterogeneity influences the practical applications such as mixing efficiency \citep{nayak2018mixing,chu2019magnetohydrodynamic,guan2021mixing}, heat and mass transfer rates \citep{ghosal2006electrokinetic,ng2012dispersion,azari2020electroosmotic} in the microfluidic devices. The `charge-heterogeneity' (CH) is defined (\eqn\ref{eq:srrh}) as the surface charge variation, {parallel} to the external pressure gradient {($\sigma \parallel\nabla P$)}, in the microfluidic device, i.e., two or more surfaces made by different materials are connected in the series.
The literature, however, includes one study \citep{xuan2008streaming} that has explored such phenomena in the electroviscous flow {by using} the phenomenological coefficients to analytically analyze the electrokinetic effects in uniform microchannel considering two types of surface charge variation perpendicular ($\sigma\perp\nabla P$) and parallel ($\sigma\parallel\nabla P$) to the external pressure gradient. The flow characteristics were noted  \citep{xuan2008streaming} to be dependent on the surface charge heterogeneity arrangement for the smaller Debye parameter ($K<50$), and such dependence becomes weak for larger $K>50$. 
On the other hand, few studies have explored the electro-osmotic flow in microfluidic devices by considering surface charge heterogeneity variation in longitudinal \citep{herr2000electroosmotic,boyko2015flow,ghosh2016electro,azari2020electroosmotic} and transverse \citep{erickson2001streaming,ng2013dispersion} directions and combining both \citep{stroock2000patterning}.

In summary,  {the literature comprises significant knowledge of electroviscous flow (EVF) in the homogeneously charged ($S_\text{rh}=1$) microfluidic devices \citep{davidson2007electroviscous,davidson2008electroviscous,bharti2008steady,bharti2009electroviscous,davidson2010electroviscous,Berry2011,dhakar2022electroviscous,dhakar2022slip,dhakar2023cfd,dhakar2024influence}; the corresponding knowledge for the heterogeneously charged ($S_\text{rh}\neq 1$) microfluidic devices is limited \citep{xuan2008streaming,dhakar2024fmfp} and suggests that surface charge-heterogeneity significantly affects microfluidic hydrodynamics. A detailed understanding of}  the electroviscous effects in heterogeneously charged ($S_\text{rh}\neq 1$) microfluidic devices remain{s} a fascinating and unexplored area of research. 

Therefore, in this work, electroviscous (EV) effects in the electrolyte liquid flow through a heterogeneously charged ($S_\text{rh}\neq 1$) uniform slit microfluidic device have been investigated numerically. Physical model governing equations such as Poisson's, Nernst-Planck (N-P), and Navier-Stokes (N-S) equations are solved numerically by using the finite element method (FEM). The effects of dimensionless flow parameters ($K$, $ {S_\text{1}}$, $ {S_\text{rh}}$) on the flow fields, such as total electrical potential ($U$), excess charge ($n^\ast$), induced electric field strength ($E_\text{x}$), pressure ($P$), and electroviscous correction factor ($Y$) are thoroughly analyzed.  Finally, a simpler pseudo-analytical model is developed to predict the pressure drop ($\Delta P$, hence, electroviscous correction factor, $Y$) in electrolyte liquid flow through a heterogeneously charged microfluidic device.
%
\section{Physical and mathematical modelling}
\begin{figure}[!hbt]
	\centering
	\includegraphics[width=1\linewidth]{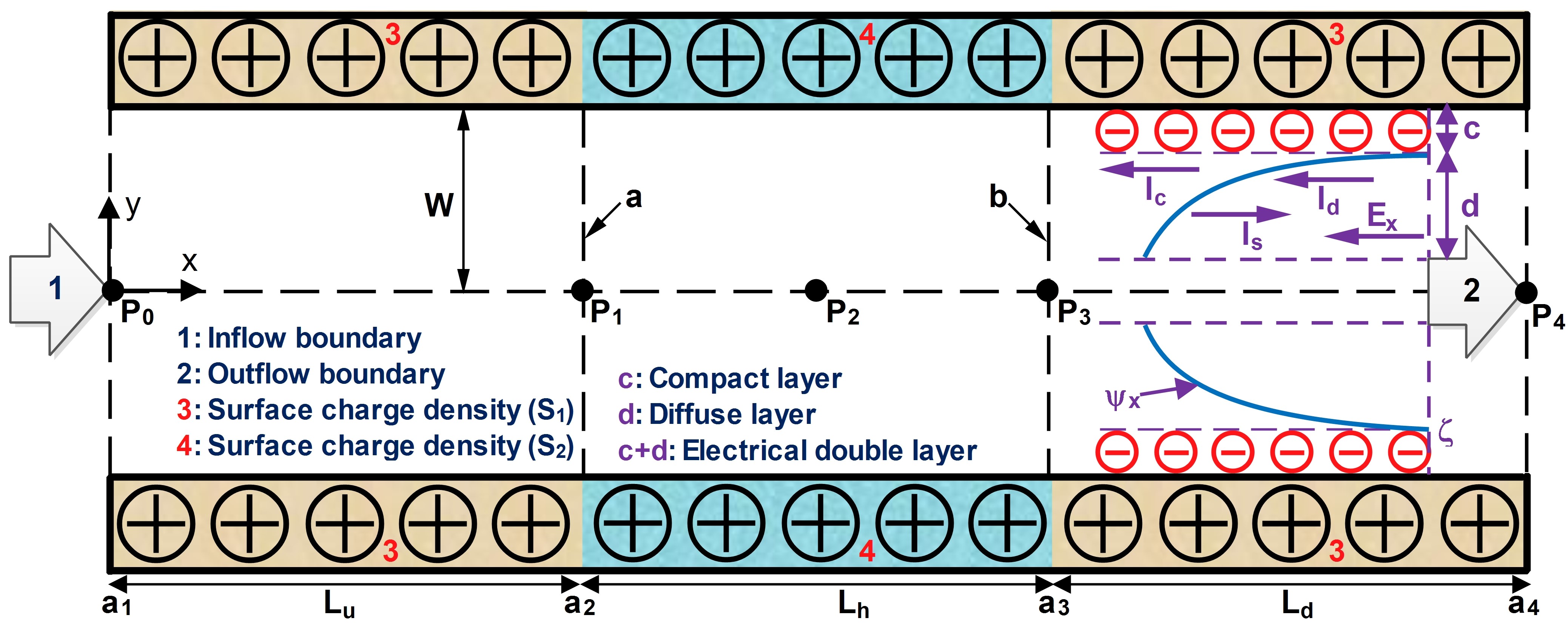}
	\caption{Schematic diagram of electroviscous flow in the heterogeneously charged slit microfluidic device.}
	\label{fig:1}
\end{figure}

%
\label{sec:mode}
Consider a steady, laminar, and fully-developed flow (volumetric flow rate, $Q$ m$^3$/s; average inflow velocity, $\overbar{V}$ m/s) of incompressible and Newtonian electrolyte liquid (density, $\rho$ kg/m$^3$; viscosity, $\mu$ Pa.s) through a two-dimensional (2-D) uniform slit microfluidic device, as depicted in \fig\ref{fig:1}. 
A symmetric (1:1) electrolyte liquid is assumed to have equal valances ($z_{+}=-z_{-}=z$) and equal diffusivity ($\mathcal{D}_{+} = \mathcal{D}_{-}=\mathcal{D}$, m$^2$/s) of ions. The geometric mean concentration of each ionic species is $n_0$ moles/m$^{3}$ \citep{harvie2012microfluidic,davidson2016numerical,dhakar2022electroviscous}. Further, the dielectric constant ($\varepsilon_{\text{r}}$) of liquid is assumed spatially uniform {and} the dielectric constant of the walls is assumed to be much smaller than liquid ($\varepsilon_{\text{r,w}} \lll \varepsilon_{\text{r}}$).

The microfluidic device of uniform cross-sectional width ($2W \mu$m) consists of three sections, i.e., upstream, heterogeneous, and downstream sections of length (in $\mu$m) $L_\text{u}$, $L_\text{h}$, and $L_\text{d}$, respectively. Thus, the total length and width of a device are $L(=L_\text{u}+L_\text{h}+L_\text{d})$ and $2W$, respectively.
Furthermore, charge-heterogeneity (CH) is considered at the device walls, i.e., both walls of the upstream and downstream sections are imposed with the surface charge density ($\sigma_{1}$, C/m$^2$), whereas the heterogeneous section walls are imposed with the surface charge density ($\sigma_{2}$, C/m$^2$), refer \fig\ref{fig:1}. While the individual walls are heterogeneously charged (i.e., $\sigma_{1}\neq \sigma_{2}$ or $S_\text{rh}\neq 1$), both walls are symmetrically charged (i.e., $\sigma_{k,t} = \sigma_{k,b}$ or $S_\text{r} = 1$).

The physical problem can be expressed by the mathematical model \citep{davidson2007electroviscous,davidson2008electroviscous,bharti2008steady,bharti2009electroviscous,davidson2010electroviscous,Berry2011,dhakar2022electroviscous,dhakar2022slip,dhakar2023cfd,dhakar2024influence} consisting of Poisson's equation (\eqn\ref{eq:1}) for total electrical potential ($U$) field, Nernst-Planck (N-P) equation (\eqn\ref{eq:5}) for ion concentration ($n_\pm$) field, Navier-Stokes (N-S) with additional body force term  (\eqns\ref{eq:9} and \ref{eq:10}) for velocity ($\myvec{V}$) and pressure ($P$) fields, respectively. 
The mathematical model (\eqns\ref{eq:1} to \ref{eq:12}) are non-dimensionalized by using the scaling factors such as $W$, $\overline{V}$, ($W/\overline{V}$), $\rho\overline{V}^2$, $U_{\text{c}} (=k_{\text{B}}T/ze)$, and $n_{\text{0}}$ for length, velocity, time, pressure, electrical potential, and the number density of ions, respectively. The dimensionless groups resulting from scaling analysis are expressed as follows.
\begin{gather}
	Re=\frac{\rho\overline{V}W}{\mu}; \quad
	 {Sc}=\frac{\mu}{\rho \mathcal{D}}; \quad
	Pe =Re\times {Sc}; \quad
	\beta=\frac{\rho\varepsilon_{\text{0}}\varepsilon_{\text{r}}U_\text{c}^2}{2\mu^2}; \quad 
	K^2=\frac{2W^2zen_{\text{0}}}{\varepsilon_{\text{0}}\varepsilon_{\text{r}} U_\text{c}}
	\label{eq:14}
\end{gather}
where $ {Sc}$, $Re$, $Pe$, $\beta$, and $K$ are the Schmidt, Reynolds, and Peclet numbers, liquid parameter, and inverse Debye length ($\lambda_\text{D}^{-1}$), respectively. Here, $\varepsilon_{\text{0}}$, $k_\text{B}$, $e$, and $T$ are the permittivity of free space, Boltzmann constant, elementary charge of a proton, and temperature, respectively. 

The dimensionless and dimensional forms of the mathematical model are presented elsewhere \citep{dhakar2022electroviscous}, and thus, to avoid duplication but maintain completeness, the dimensionless form of the governing equations and relevant boundary conditions (BC) for each flow field is subsequently expressed as follows (retaining the variable names the same as in dimensional form). 
\subsection{Electrical potential field}
\noindent 
The distribution of the total electrical potential ($U$) field in the microfluidic device can be described by Poisson's equation (\eqn\ref{eq:1}) relating the total potential ($U$) with a  {local} charge density of ions ($\rho_{\text{e}}$) \citep{davidson2007electroviscous,bharti2008steady,dhakar2022electroviscous,dhakar2023cfd,dhakar2024influence,davidson2010electroviscous,Berry2011} as follows.
\begin{gather}
	\nabla^2U=-\frac{1}{2}K^2\rho_{\text{e}}
	\label{eq:1}
\end{gather}
where $\rho_{\text{e}} \equiv n^\ast(=n_+-n_-)$ is the excess charge for symmetric electrolyte, and $n_\text{j}$ is the number density of $j^{th}$ ion, respectively.
In general, total electrical potential ($U$) in electroviscous flow (EVF) is expressed  \citep{davidson2007electroviscous,bharti2008steady,bharti2009electroviscous,dhakar2022electroviscous,dhakar2023cfd,dhakar2024influence,davidson2010electroviscous,Berry2011} as follows.
\begin{gather}
	U(x,y)=\psi(y)-\phi(x)
	\label{eq:4a}
\end{gather}
where $\phi~(=xE_\text{x})$, $\psi$, $E_\text{x}$, $x$, and $y$ are the streaming and EDL potentials, induced electric field strength in the axial flow direction, axial and transverse coordinates, respectively. 
{Since streaming potential ($\phi$) varies linearly along the device in a homogeneously charged ($S_\text{rh}=1$) uniform microchannels, EDL and streaming potentials can be decoupled \citep{bharti2009electroviscous}. However, such a decoupling is not possible for heterogeneously charged ($S_\text{rh}\neq1$) devices due to non-linear variation of the streaming potential ($\phi$) along the channel \citep{davidson2007electroviscous,bharti2008steady,dhakar2022electroviscous,dhakar2023cfd,dhakar2024influence,davidson2010electroviscous,Berry2011}, and thus, the total potential ($U$) needs to be analyzed. }
The potential field (\eqn\ref{eq:1}) is subjected to the following boundary conditions (BC).

Uniform potential gradient is applied at inlet ($x=0$) and outlet ($x=L$) boundaries, which is obtained by satisfying \citep{davidson2007electroviscous,dhakar2022electroviscous,dhakar2023cfd,dhakar2024influence} the \textit{current continuity condition} ($I_\text{net}=\nabla\cdot I=0$)  
written as follows.
\begin{gather}
	I_{\text{net}} = \underbrace{\int_{-1}^{1} {n^\ast\myvec{V}} dy} _{I_{\text{s}}} - \underbrace{\int_{-1}^{1} {Pe^{-1}\left[\frac{\partial n_{\text{+}}}{\partial x}-\frac{\partial n_{\text{-}}}{\partial x}\right]} dy}_{I_{\text{d}}} - \underbrace{\int_{-1}^{1} {Pe^{-1}\left[(n_{\text{+}}+n_{\text{-}})\frac{\partial U}{\partial x}\right]} dy}_{I_{\text{c}}} =0
	\label{eq:2}
\end{gather}
where $\myvec{V}$, $I_\text{s}$, $I_\text{c}$, and $I_\text{d}$ are the velocity field (\eqn\ref{eq:9}), streaming, conduction, and diffusion currents, respectively. At steady-state, the diffusion current is zero (i.e., $I_\text{d}=0$). 

The symmetrically 
positively charged ($S_\text{r} = 1$, \eqn\ref{eq:srrh}) walls ($y=\pm W$) are imposed with the charge-heterogeneity 
(CH)  as follows.
\begin{gather}
	(\nabla U\cdot\myvec{n}_{b}) = 
	\begin{cases}	
		{S_{2} \ge 0} \quad \text{for} \ a_2 < x < a_3 
		\\ 
		{S_{1} > 0} \quad \text{otherwise} 
	\end{cases}\label{eq:3}
\end{gather}
{Refer \fig\ref{fig:1} for $a_2$ and $a_3$. In this study, the} surface charge-heterogeneity ratio 
(${S_\text{rh}}$, \eqn\ref{eq:srrh}) is {written} as follows.
\begin{gather}
	{S_\text{rh}}=\frac{S_{2}}{S_{1}} \qquad
		\text{where}\qquad
	{S_{\text{k}}}=\frac{\sigma_{\text{k}}W}{\varepsilon_{\text{0}}\varepsilon_\text{r} U_\text{c}}, \qquad
	k=1, 2
	\label{eq:4}
\end{gather}
where $S_{\text{k}}$, and $\myvec{n}_{b}$ are the dimensionless surface charge density of the k$^\text{th}$ section of the wall, and unit vector normal to the wall, respectively.
%
Note that, in the case of ${S_\text{rh}}=0$, only upstream and downstream sections walls are charged (${S_{1}} > 0$), and heterogeneous section walls are electrically neutral (i.e., ${S_{2}}=0$); and each section of microfluidic device is homogeneously charged for ${S_\text{rh}}=1$. The upstream and downstream sections walls charge (${S_{1}}$) dominates for ${S_\text{rh}}<1$, and heterogeneous section walls charge (${S_{2}}$) dominates for ${S_\text{rh}}>1$. Further, the non-electroviscous flow (nEVF) condition can be imposed by setting ${S_{\text{k}}}=0$ in \eqn (\ref{eq:3}) on the walls.
\subsection{Ion concentration field}
%
The distribution of ion concentration ($n_\pm$) field in the microfluidic device can be described by the Nernst-Planck (N-P) equation (\eqn\ref{eq:5}) depicting the conservation of each ${\text{j}}^{\text{th}}$ ionic species \citep{davidson2007electroviscous,bharti2008steady,dhakar2022electroviscous,dhakar2023cfd,dhakar2024influence} as follows.
\begin{gather}
	\left[\frac{\partial n_{\text{j}}}{\partial t}+\nabla\cdot(\myvec{V}n_{\text{j}})\right]=(1/Pe)\left[\nabla^2n_{\text{j}}\pm\nabla\cdot(n_{\text{j}}\nabla U)\right]
	\label{eq:5}
\end{gather}
where $t$ is time. 
{\eqn(\ref{eq:5})}  is subjected to the following boundary conditions (BC).
%
The ion concentration ($n_{\text{j}}$) field obtained from the numerical solution of steady fully developed electroviscous (EV) flow through uniform micro-slit \citep{davidson2007electroviscous,bharti2008steady,bharti2009electroviscous,dhakar2024influence}
is imposed at inlet ($x=0$) boundary. 
An ionic concentration gradient at the outlet ($x=L$) and the flux density of ions normal to the walls ($y=\pm W$) are applied as zero. 
\begin{gather}
	n_\pm=\exp[{\mp\psi(y)}],\qquad  \text{at inlet  ($x=0$)}
	\label{eq:6}  \\
%
	\frac{\partial n_{\text{j}}}{\partial x} = 0,  \qquad\qquad \text{at outlet  ($x=L$)} \label{eq:7a}\\
	\myvec{f}_{\text{j}}\cdot \myvec{n}_{\text{b}}=0,   \qquad \text{at walls  ($y=\pm W$)}
	\label{eq:7}
\end{gather}
where $\myvec{f}_{\text{j}}$ is flux density of ${\text{j}}^{\text{th}}$ species defined by the Einstein relation \citep{dhakar2022electroviscous}.
\subsection{Flow field}
The distribution of flow velocity ($\myvec{V}$) and pressure ($P$) fields in the microfluidic device can be described by the Navier-Stokes (N-S)  {equations, i.e., momentum conservation equation} with electrical body force (\eqn\ref{eq:9}), and mass conversation equations (\eqn\ref{eq:10}) for incompressible electrolyte liquid flow \citep{davidson2007electroviscous,dhakar2022electroviscous,dhakar2023cfd,dhakar2024influence} as follows. 
\begin{gather}
	\left[\frac{\partial \mathbf{V}}{\partial t}+\nabla\cdot(\myvec{V}\myvec{V})\right]=-\nabla P+(1/Re)\nabla \cdot\left[\nabla\myvec{V}+(\nabla\myvec{V})^T\right]-\underbrace{\beta (K/Re)^2n^\ast\nabla U}_{\myvec{F}_{\text{e}}}
	\label{eq:9}
	\\
	\nabla\cdot\myvec{V}=0 \label{eq:10}
\end{gather}
where $\myvec{F}_{\text{e}}$ is the electrical body force. 
The flow field equations (\eqns\ref{eq:9} and \ref{eq:10}) are subjected to the following boundary conditions (BC).
A fully developed velocity field, $V_{0}(y)$, obtained from the numerical solution of steady electroviscous (EV) flow through uniform micro-slit  \citep{davidson2007electroviscous,bharti2008steady,bharti2009electroviscous,dhakar2024influence} is applied at the inlet ($x=0$) boundary. 
%
%
%
The velocity gradient is applied to be zero at the outlet ($x=L$) boundary open to the ambient (i.e., atmosphere). No-slip velocity condition is applied at device walls ($y=\pm W$).  Mathematically, 
%
\begin{align}
		V_{\text{x}} =V_{\text{0}}(y), & & V_{\text{y}} =0  & & \text{at inlet  ($x=0$)}& 	\label{eq:11}\\
		\frac{\partial \myvec{V}}{\partial x} = 0, & & P =0  & &\text{at outlet  ($x=L$)}& \label{eq:12a}\\
		V_{\text{x}} =0, & & V_{\text{y}} =0  & & \text{at walls  ($y=\pm W$)}&
	\label{eq:12}
\end{align}
where, $V_\text{x}$ and $V_\text{y}$ are the velocity components in $x-$ and $y-$ directions, respectively. 

The mathematical model equations with relevant boundary conditions (\eqns\ref{eq:1} to \ref{eq:12}) are solved using a finite element method (FEM) to obtain the flow fields such as total electrical potential ($U$), ion concentration ($n_\pm$), excess charge ($n^\ast$), velocity ($\myvec{V}$) and pressure ($P$) fields over the wide range of conditions ($K$, $S_{1}$, $S_{\text{rh}}$) in the heterogeneously charged microfluidic device. 
%
\section{Numerical approach}
\label{sec:sanp}
%
\noindent 
The detailed numerical approach has been described in the recent studies \citep{dhakar2022electroviscous,dhakar2023cfd,dhakar2024influence} and thus, only essential features are presented here. In this work, a finite element method (FEM) based {commercial} computational fluid dynamics (CFD) solver, COMSOL Multiphysics software, has been used to solve the {steady-state form of the} mathematical model (\eqns\ref{eq:1} to \ref{eq:12}), represented by \textit{electrostatic} (es), \textit{transport of dilute species} (tds), and \textit{laminar flow} (spf) COMSOL modules, depicting the electroviscous flow in a heterogeneously charged slit microfluidic device. An \textit{intop} function described in the global function definition section of COMSOL model coupling is used to evaluate the integral quantities in \eqn(\ref{eq:2}). The discretized set of equations has been solved iteratively using the fully coupled PARDISO linear and Newton's non-linear solvers {with the convergence criteria of $10^{-5}$}. The steady-state numerical solution yields 
electroviscous flow fields ($U$, $n^\ast$, $\myvec{V}$, $P$, $E_\text{x}$, $Y$) as a function of flow governing parameters ($K$, ${S_\text{1}}$, ${S_\text{rh}}$). 

{Furthermore, to obtain the results free from mesh artifacts, the mesh independence test is carried out using three different meshes (M1, M2, and M3; \tab\ref{tab:1m}) for extreme values of flow parameters ($K=2, 20$; ${S_\text{1}}=4, 16$ ${S_\text{rh}}=1$).  The mesh is characterized by the number of uniformly distributed mesh points per unit dimensionless length ($\Delta$), the total number of mesh elements ($Ne$), and the degree of freedom (DoF). In addition to the mesh details, \tab\ref{tab:1m} includes the pressure drop ($|\Delta P^{\ast}| = 10^{-3}|\Delta P|$) values and their relative change ($\delta$, \%) from coarse to fine mesh. The mesh test results depict that the pressure drop ($|\Delta P^{\ast}|$) changes maximally by 0.03\% with the mesh variation from M1 to M3. Thus, M2 mesh is used to discretize the flow domain to obtain the final results free from mesh effects presented hereafter.}
{\begin{table}[h]
	\centering
	\caption{Mesh characteristics and influence of mesh on the pressure drop ($|\Delta P^{\ast}| = 10^{-3}|\Delta P|$) in homogeneously charged ($S_\text{rh}=1$) uniform microfluidic device.}\label{tab:1m}
\scalebox{0.95}
{\renewcommand{\arraystretch}{1.5}	
\begin{tabular}{|c|c|c|c|c|c|c|c|}\hline
		\multicolumn{4}{|c|}{Mesh details} &	\multicolumn{2}{c}{$|\Delta P^{\ast}|$ at $S_{1}=4$}	&	\multicolumn{2}{|c|}{$|\Delta P^{\ast}|$ at $S_{1}=16$}		\\	\hline
	&	$\Delta$	&	$N_\text{e}$	&	DoF	&	$K=2$	&	$K=20$	&	$K=2$	&	$K=20$	\\	\hline
	M1	&	50	&	130000	&	1179484	&	5.3214	&	4.4930	&	6.3221	&	4.5007	\\	
	M2	&	100	&	408000	&	3687434		&	5.3258	&	4.4968	&	6.3253	&	4.5047	\\	
	M3	&	150	&	1140000	&	10288384	&	5.3273	&	4.4980	&	6.3265	&	4.5059	\\	\hline
	\multicolumn{4}{|r|}{$\delta(|\Delta P^{\ast}|)_\text{M1 - M2}$, \%}	&	0.08	& 0.08	&	0.05 &	0.09	\\	
	\multicolumn{4}{|r|}{$\delta(|\Delta P^{\ast}|)_\text{M1 - M3}$, \%}	& 0.11		&	0.11 &	0.07 & 0.12	\\	
	\multicolumn{4}{|r|}{$\delta(|\Delta P^{\ast}|)_\text{M2 - M3}$, \%}	&	0.03	&	0.03 &	0.02 & 0.03		\\	
	\hline
\end{tabular}}
\end{table}}

Based on our previous experience on detailed domain and mesh independence studies and existing knowledge \citep{davidson2007electroviscous,bharti2008steady,bharti2009electroviscous,dhakar2022electroviscous,dhakar2023cfd,dhakar2024influence,Berry2011,davidson2010electroviscous}, the following numerical parameters are adopted in the simulations: (i) \textit{Geometrical parameters}: $W=0.1~\mu$m; $L_\text{u}=L_\text{h}=L_\text{d}=5W$; (ii) \textit{Mesh characteristics}: uniform, rectangular, structured mesh: M2, uniformly distributed grid points per unit length of device, $\Delta=100$; total number of mesh elements, $N_\text{e}=408000$; degree of freedom, DoF $=3687434$  (refer \tab\ref{tab:1m}). Subsequently, the new results, free from ends and mesh effects, are presented and discussed in the next section.
%
\begin{table}[!bt]
	\centering
	\caption{Parameters considered in the present study. The `EVF' and `nEVF' represent the electroviscous and non-electroviscous flows.}\label{tab:1b}
	\scalebox{0.95}
	{\renewcommand{\arraystretch}{1.5}
		\begin{tabular}{|l|c|c|c|c|}
			\hline
			Parameters & $K$&${S_\text{1}}$&${S_\text{rh}}$& Fixed \\\hline
			Range & $2 - 20$, $\infty$ &0, $4 - 16$&$0 - 2$& ${S_\text{r}=1}$, $Re = 10^{-2}$ \\ \cline{1-4}
			Values (for EVF) &$2K~|~K\in[1..10]$&$2{S_\text{1}}~|~{S_\text{1}}\in[2^1..2^3]$&$0.25{S_\text{rh}}~|~{S_\text{rh}}\in[0..8]$  & $ {Sc}=10^3$  \\ \cline{1-4}
			Values (for nEVF)&\multicolumn{3}{c|}{$K=\infty$ or ${S_\text{k}}=0$} & $\beta = 2.34\times10^{-4}$\\ \hline
		\end{tabular}
	}	
\end{table}
\section{Results and discussion}
\noindent 
In the study, systematic parametric investigation has been performed to analyze the electroviscous effects in the pressure-driven flow of symmetric electrolyte through heterogeneously charged uniform slit microfluidic device for the wide range of the conditions listed in \tab\ref{tab:1b}. The justification of the considered ranges of parameters ($ {Sc}$, $Re$, $\beta$, $ {S_\text{1}}$, $K$; \tab\ref{tab:1b}) have been presented in the recent literature \citep{davidson2007electroviscous,dhakar2022electroviscous, dhakar2023cfd,dhakar2024influence}, and the present modeling approach has also thoroughly been validated previously \citep{dhakar2022electroviscous, dhakar2023cfd}.  Therefore, this section has presented the new detailed numerical results in terms of flow fields such as total potential ($U$), ion concentration ($n_\pm$), excess charge ($n^\ast$), induced electric field strength ($E_\text{x}$), pressure ($P$), and electroviscous correction factor ($Y$) for given ranges of parameters (\tab\ref{tab:1b}). %
In addition, the relative impact of charge-heterogeneity (${S_\text{rh}}$) on the flow fields is analyzed by normalizing them for heterogeneously charged (${S_\text{rh}}\neq1$) by that at reference case of homogeneously charged (`ref' or $ {S_\text{rh}}=1$) device for {similar values of the} dimensionless parameters ($ {S_\text{1}}$, $K$) as defined below.
\begin{gather}
	\Psi_\text{n} = \frac{\Psi}{\Psi_{\text{ref}}} = 
	\left.\frac{\Psi( {S_\text{rh}})}{\Psi( {S_\text{rh}}=1)}\right|_{ {S_\text{1}}, K}
	\qquad\text{where}\qquad \Psi = (\Delta U, n^{\ast}, E_{\text{x}}, \Delta P)
	\label{eq:11a}
\end{gather}
where, $\Delta U$ and $\Delta P$ are the electrical potential and pressure drops, respectively.
%
\subsection{Total electrical potential ($U$) distribution}
\label{sec:potential}
\begin{figure}[!b]
	\centering\includegraphics[width=1\linewidth]{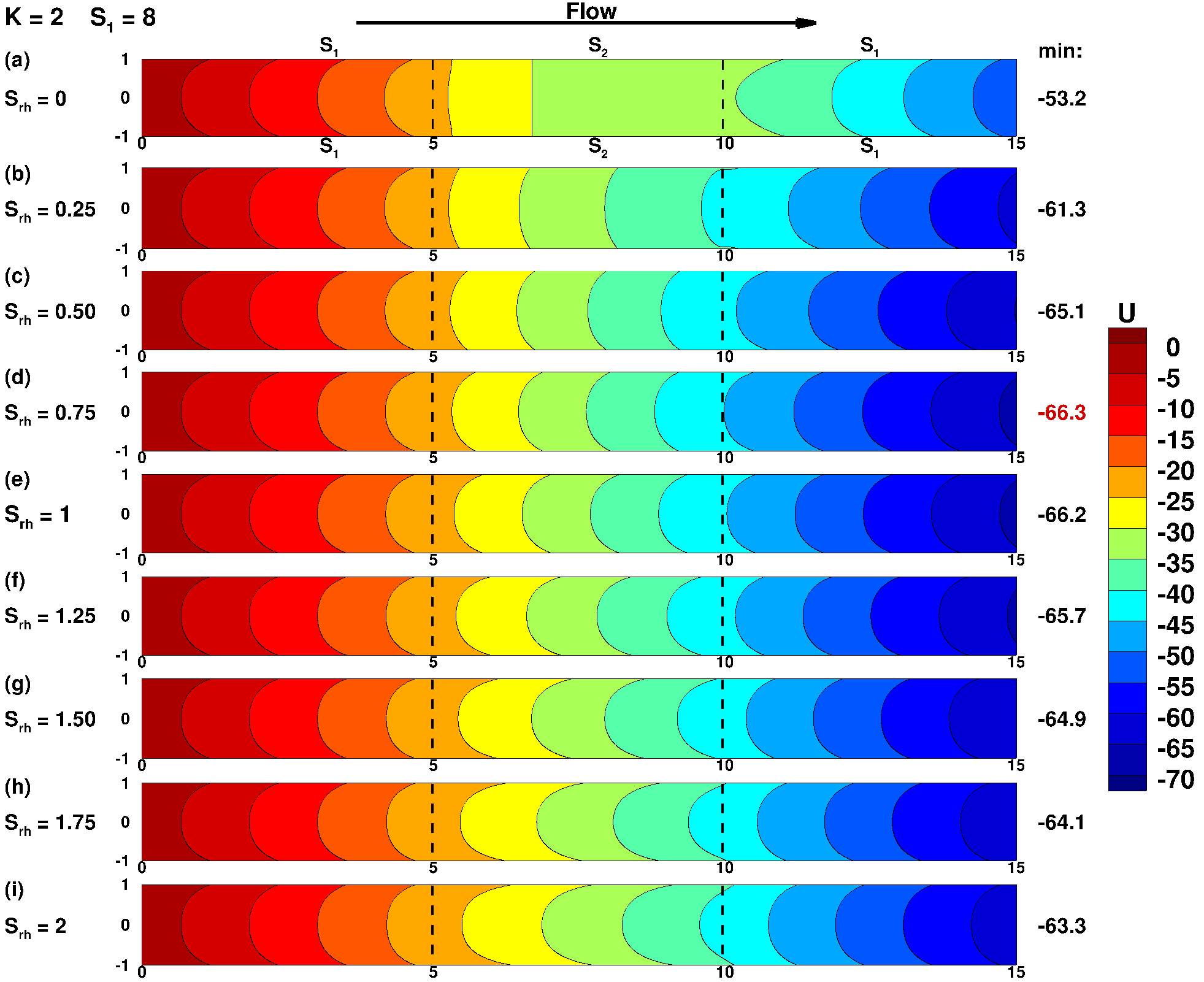}
	\caption{Influence of charge heterogeneity ($0\le S_\text{rh}\le 2$) on the total electrical potential ($U$) for fixed conditions ($S_\text{1}=8$ and $K=2$).}
	\label{fig:2}
\end{figure}

%
In electroviscous flow (EVF), the distribution of the total electrical potential ($U$) in the homogeneously charged (${S}_\text{rh}=1$)  microfluidic devices is known \citep{davidson2007electroviscous,dhakar2022electroviscous, dhakar2023cfd,dhakar2024influence} to strongly depend on the flow parameters ($K$, ${S_\text{k}}$). For heterogeneously charged (${S}_\text{rh}\neq 1$)  uniform microfluidic device, \fig\ref{fig:2} depicts an influence of charge heterogeneity ($0\le S_\text{rh}\le 2$) on the total potential ($U$) contours for fixed conditions ($S_\text{1}=8$ and $K=2$); qualitatively similar profiles are observed for other conditions (\tab\ref{tab:1b}).
%
Broadly, in a positively charged (${S_\text{k}} > 0$) device,  the potential reduces along the length ($0\le x\le L$) due to the advection of negative ions (\fig\ref{fig:2}). The lateral curving of the contours is seen due to fixed potential gradient ($\partial U/\partial\myvec{n}_{\text{b}} = S_\text{k} \neq0$) at device walls (except heterogeneous section in \fig\ref{fig:2}a). Further, the contour profiles are observed symmetric about the centreline ($P_0$ to $P_4$; \fig\ref{fig:1}) for homogeneously charged ($S_\text{rh}=1$) microchannel. 
The shape of contours is remarkably affected in the heterogeneous section of the device, i.e., it changes from uniform to convex shape with increasing $S_\text{rh}$ from 0 to 2 at fixed $K$ and $S_\text{1}$ (\fig\ref{fig:2}). 
However, contours are relatively less affected in the downstream region with enhancing $ {S_\text{rh}}$ followed by negligibly affected in upstream than heterogeneous section of the device, irrespective of $K$ and $S_\text{1}$. For $S_\text{rh}<1$, at line $a$ ($x=5$; \fig\ref{fig:1}), excess charge moves in the axial flow direction, i.e., from upstream to heterogeneous section whereas at line $b$ ($x=10$; \fig\ref{fig:1}), excess charge {advects} opposite to the flow direction due to varied charge gradient. However, the flow of excess charge at lines $a$ and $b$ has shown reverse trends for $S_\text{rh}>1$ for given ranges of conditions. Therefore, sudden change in the shape of contours obtained near the points of charge heterogeneity (lines $a$ and $b$) in the device (\fig\ref{fig:2}). 

Further, the potential ($U$) decreases with increasing $S_\text{rh}$ (\eqn\ref{eq:4}) followed by reverse trends at higher $S_\text{rh}$ (\fig\ref{fig:2}) due to strengthening the charge-attractive force close to channel walls which  {accumulates} excess charge in the EDL. Thus, the streaming current ($I_{\text{s}}$) enhances, and total potential reduces with increasing $S_\text{rh}$, but at higher $S_\text{rh}$, electrostatic force {is} remarkably stronger, which impedes excess ions flow in the microfluidic device. Further, overall minimum potential ($U_{\text{min}}$) is noted as $-66.30$ at $S_\text{rh}=0.75$ for $K=2$ and $S_\text{1}=8$ (\fig\ref{fig:2}d). 
\begin{figure}[!b]
	\centering\includegraphics[width=1\linewidth]{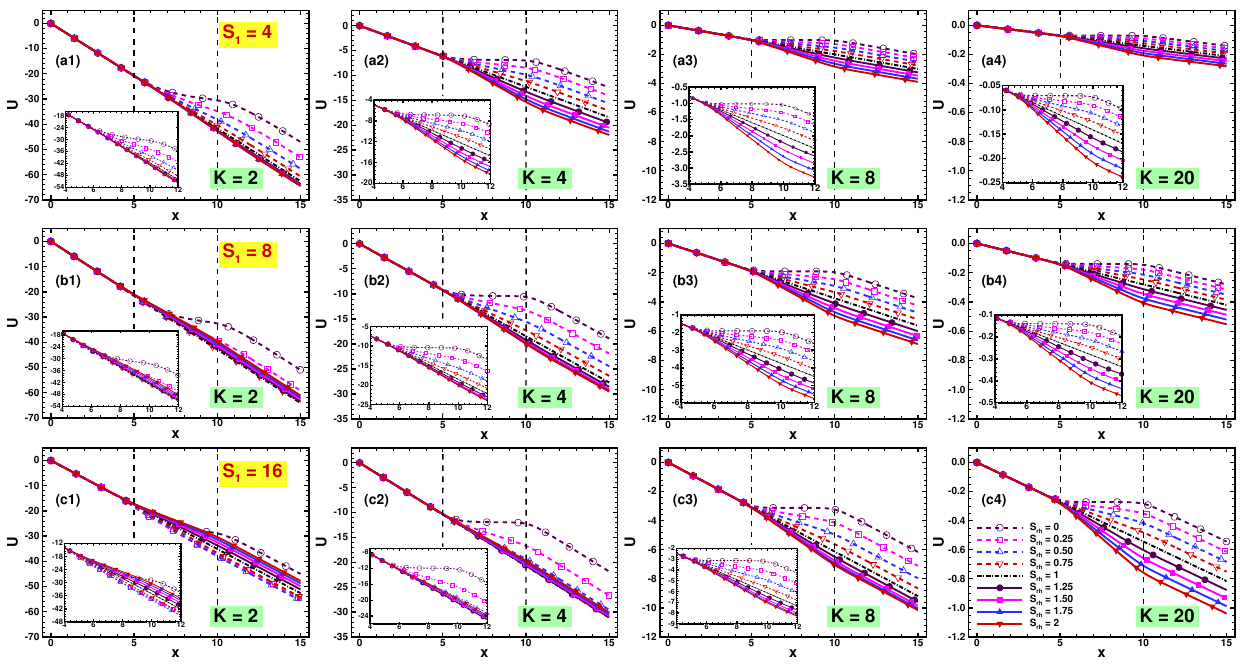}
	\caption{Total potential ($U$) variation on the centreline ($P_0$ to $P_4$; \fig\ref{fig:1}) of heterogeneously charged microfluidic device for dimensionless parameters ($K$, $ {S_\text{1}}$, $ {S_\text{rh}}$; \tab\ref{tab:1b}).}
	\label{fig:3}
\end{figure} 
%

Subsequently, \fig\ref{fig:3} depicts the total potential ($U$) variation on the centreline ($P_0$ to $P_4$; \fig\ref{fig:1}) over the considered ranges of conditions (\tab\ref{tab:1b}). In general, the potential decreases along the length ($0\le x\le L$) of the channel. In upstream ($0\le x\le 5$) and downstream ($10\le x\le 15$) sections, the potential gradient varies linearly (or uniformly) along the length, irrespective of the flow conditions. However, in the heterogeneous ($5\le x\le 10$) section, it changes non-linearly from minimum to maximum (\fig\ref{fig:3}) with enhancing charge-heterogeneity ($0\le S_\text{rh}\le 2$), irrespective of other flow conditions ($K$, $S_\text{1}$). It is due to increased clustering of the available excess charge for transport in the heterogeneous section with the enhancement of $S_\text{rh}$ at fixed $K$ and $S_\text{1}$. 
The electrical potential ($U$) has shown qualitatively similar dependence \citep{davidson2007electroviscous,dhakar2022electroviscous, dhakar2023cfd,dhakar2024influence} as that for the homogeneously charged ($S_\text{rh}=1$) walls on $K$ and $S_\text{1}$, i.e., it decreases with decreasing $K$ or thickening of EDL in the microfluidic device (\fig\ref{fig:3}), as the decreasing K leads to the thickening of EDL, which provides an increasing large cross-section of channel for the interaction of ions with charged surface and thus reducing electrical potential. Further, $U$ has shown complex dependency on $S_\text{1}$ and $S_\text{rh}$. For instance, total potential reduces with increasing $S_\text{1}$ and $S_\text{rh}$ followed by reverse trends at higher $S_\text{1}$ and $S_\text{rh}$ (\fig\ref{fig:3}). The maximum variation in the potential drop ($|\Delta U|$) is recorded as 287.20\% {({from} 0.1766 to 0.6838)} at $S_\text{rh}=0.50$ and $K=20$ with increasing $S_\text{1}$ from 4 to 16 (refer \fig\ref{fig:3}).
%

%
\begin{sidewaystable}	
	\centering
	\caption{Total potential drop ($|\Delta U|$), critical excess charge ($n^\ast_{\text{c}}$), critical induced electric field strength ($E_{\text{x,c}}$), and pressure drop ($|\Delta P^{\ast}|$) on the centreline ($P_0$ to $P_4$; \fig\ref{fig:1}) of heterogeneously charged microfluidic device. The critical values ($n^\ast_{\text{c}}$, $E_{\text{x,c}}$) are noted as maximum (superscript $\varoplus$) and minimum (superscript $\circleddash$).}\label{tab:1}
	\scalebox{0.49}
	{\renewcommand{\arraystretch}{1.5}
		\begin{tabular}{|r|r|r|r|r|r|r|r|r|r|r|r|r|r|r|r|r|r|r|r|r|}
			\hline
			$ {S_\text{1}}$	&	$K$	&	\multicolumn{9}{c|}{$|\Delta U|$} &	\multicolumn{10}{c|}{$n^\ast_{\text{c}}\times10^{\alpha}$}	\\\cline{3-21}
			&		&	$ {S_\text{rh}}=0$	&	$ {S_\text{rh}}=0.25$	& $ {S_\text{rh}}=0.50$ &	$ {S_\text{rh}}=0.75$	& $ {S_\text{rh}}=1$ & $ {S_\text{rh}}=1.25$ & $ {S_\text{rh}}=1.50$ & $ {S_\text{rh}}=1.75$ & $ {S_\text{rh}}=2$ &	$ {S_\text{rh}}=0^\varoplus$	&	$ {S_\text{rh}}=0.25^\varoplus$	& $ {S_\text{rh}}=0.50^\varoplus$ &	$ {S_\text{rh}}=0.75^\varoplus$	& $ {S_\text{rh}}=1^{\circleddash}$ & $ {S_\text{rh}}=1.25^{\circleddash}$ & $ {S_\text{rh}}=1.50^{\circleddash}$ & $ {S_\text{rh}}=1.75^{\circleddash}$ & $ {S_\text{rh}}=2^{\circleddash}$ & $\alpha$  \\\hline
			\multicolumn{2}{|c|}{nEVF}    & 	0	    & 0	        & 0          & 	0	    & 0         & 	0	& 0	&0	        &0 & 	0	    & 0	        & 0          & 	0	    & 0         & 	0	& 0	&0	        &0 & 0 \\\hline
			4	&	2	&	47.0420	&	53.1350	&	57.6670	&	60.7010	&	62.5850	&	63.6750	&	64.2420	&	64.4670	&	\textbf{64.4670}	&	-1.4528	&	-3.4167	&	-5.2251	&	-6.8129	&	-8.1585	&	-9.2814	&	-10.2170	&	-11.0010	&	\textbf{-11.6630}	&	1	\\
			&	4	&	12.3910	&	14.0610	&	15.6540	&	17.1180	&	18.4210	&	19.5510	&	20.5080	&	21.3070	&	\textbf{21.9630}	&	-0.0807	&	-1.8324	&	-3.5315	&	-5.1372	&	-6.6217	&	-7.9708	&	-9.1819	&	-10.2600	&	\textbf{-11.2160}	&	2	\\
			&	6	&	4.4124	&	4.9820	&	5.5425	&	6.0862	&	6.6066	&	7.0984	&	7.5578	&	7.9827	&	\textbf{8.3718}	&	-0.0026	&	-1.6224	&	-3.2243	&	-4.7918	&	-6.3105	&	-7.7689	&	-9.1582	&	-10.4730	&	\textbf{-11.7100}	&	3	\\
			&	8	&	2.0007	&	2.2549	&	2.5072	&	2.7556	&	2.9985	&	3.2344	&	3.4621	&	3.6805	&	\textbf{3.8887}	&	-0.0001	&	-1.6520	&	-3.2945	&	-4.9186	&	-6.5161	&	-8.0796	&	-9.6029	&	-11.0810	&	\textbf{-12.5090}	&	4	\\
			&	20	&	0.1413	&	0.1589	&	0.1766	&	0.1942	&	0.2118	&	0.2293	&	0.2467	&	0.2640	&	\textbf{0.2812}	&	0.0414	&	-0.3757	&	-0.7089	&	-1.1251	&	-1.6119	&	-2.1114	&	-2.7777	&	-3.5973	&	\textbf{-4.5937}	&	9	\\\hline
			8	&	2	&	51.0220	&	59.1880	&	62.9600	&	\textbf{64.1140}	&	64.0740	&	63.5250	&	62.7740	&	61.9630	&	61.1550	&	-0.1688	&	-0.5267	&	-0.8037	&	-1.0001	&	-1.1391	&	-1.2402	&	-1.3160	&	-1.3746	&	\textbf{-1.4209}	&	0	\\
			&	4	&	18.9280	&	21.9370	&	24.4760	&	26.3880	&	27.7130	&	28.5740	&	29.0980	&	29.3890	&	\textbf{29.5180}	&	-0.0118	&	-0.3386	&	-0.6322	&	-0.8768	&	-1.0724	&	-1.2264	&	-1.3476	&	-1.4438	&	\textbf{-1.5208}	&	1	\\
			&	6	&	7.8189	&	8.9195	&	9.9544	&	10.8790	&	11.6700	&	12.3230	&	12.8490	&	13.2620	&	\textbf{13.5810}	&	-0.0004	&	-0.3114	&	-0.6095	&	-0.8848	&	-1.1317	&	-1.3488	&	-1.5371	&	-1.6993	&	\textbf{-1.8385}	&	2	\\
			&	8	&	3.7529	&	4.2538	&	4.7397	&	5.1980	&	5.6197	&	5.9994	&	6.3353	&	6.6282	&	\textbf{6.8805}	&	-0.0001	&	-3.2221	&	-6.3721	&	-9.3900	&	-12.2310	&	-14.8690	&	-17.2910	&	-19.4970	&	\textbf{-21.4940}	&	4	\\
			&	20	&	0.2803	&	0.3156	&	0.3508	&	0.3857	&	0.4202	&	0.4543	&	0.4877	&	0.5204	&	\textbf{0.5524}	&	2.2936	&	1.8428	&	0.8270	&	-0.8049	&	-3.1843	&	-6.1562	&	-10.0310	&	-14.8470	&	\textbf{-20.7170}	&	9	\\\hline
			16	&	2	&	44.9890	&	54.6570	&	\textbf{55.5480}	&	54.2460	&	52.6360	&	51.1370	&	49.8310	&	48.7080	&	47.7470	&	-0.1684	&	-0.7646	&	-1.0839	&	-1.2539	&	-1.3549	&	-1.4202	&	-1.4656	&	-1.4988	&	\textbf{-1.5241}	&	0	\\
			&	4	&	21.8450	&	26.6960	&	29.5530	&	30.7740	&	\textbf{31.1170}	&	31.0400	&	30.7760	&	30.4350	&	30.0720	&	-0.0126	&	-0.5719	&	-0.9768	&	-1.2352	&	-1.3999	&	-1.5086	&	-1.5835	&	-1.6368	&	\textbf{-1.6758}	&	1	\\
			&	6	&	11.2970	&	13.3020	&	14.9150	&	16.0260	&	16.7180	&	17.1100	&	17.3080	&	\textbf{17.3800}	&	17.3740	&	-0.0006	&	-0.5591	&	-1.0404	&	-1.4169	&	-1.6990	&	-1.9083	&	-2.0652	&	-2.1847	&	\textbf{-2.2772}	&	2	\\
			&	8	&	6.1526	&	7.1095	&	7.9630	&	8.6584	&	9.1889	&	9.5757	&	9.8479	&	10.0330	&	\textbf{10.1530}	&	-0.0002	&	-5.9765	&	-11.4730	&	-16.2260	&	-20.1850	&	-23.4170	&	-26.0370	&	-28.1850	&	\textbf{-29.9660}	&	4	\\
			&	20	&	0.5441	&	0.6145	&	0.6838	&	0.7512	&	0.8158	&	0.8771	&	0.9347	&	0.9884	&	\textbf{1.0382}	&	3.3377	&	2.9920	&	2.2211	&	1.0195	&	-0.6127	&	-2.6028	&	-4.9555	&	-7.6140	&	\textbf{-10.5360}	&	8	\\\hline
		%
			${S_\text{1}}$	&	$K$	&	\multicolumn{9}{c|}{$E_{\text{x,c}}$ (or $^{\dagger}E_{\text{x,c}}\times10^{\alpha}$)} &	\multicolumn{9}{c|}{$|\Delta P^{\ast}| =|\Delta P|\times 10^{-3}$} &	\\\cline{3-20}
			&		&	${S_\text{rh}}=0^{\circleddash}$	&	${S_\text{rh}}=0.25^{\circleddash}$	& ${S_\text{rh}}=0.50^{\circleddash}$ &	${S_\text{rh}}=0.75^{\circleddash}$	& ${S_\text{rh}}=1^\varoplus$ & $ {S_\text{rh}}=1.25^\varoplus$ & ${S_\text{rh}}=1.50^\varoplus$ & ${S_\text{rh}}=1.75^\varoplus$ & ${S_\text{rh}}=2^\varoplus$ &	$ {S_\text{rh}}=0$	&	$ {S_\text{rh}}=0.25$	& ${S_\text{rh}}=0.50$ &	${S_\text{rh}}=0.75$	& ${S_\text{rh}}=1$ & ${S_\text{rh}}=1.25$ & ${S_\text{rh}}=1.50$ & $ {S_\text{rh}}=1.75$ & ${S_\text{rh}}=2$ &  $\alpha$ \\\hline
			\multicolumn{2}{|c|}{nEVF}	&	0	&	0	&	0	&	0	&	0	&	0	&	0	&	0	&	0	&	4.4962	&	4.4962	&	4.4962	&	4.4962	&	4.4962	&	4.4962	&	4.4962	&	4.4962	&	4.4962&	-- \\\hline
			4	&	2	&	0.7542	&	2.1299	&	3.1488	&	3.7962	&	4.1720	&	4.4021	&	4.5548	&	4.6572	&	\textbf{4.7252}	&	4.9971	&	5.0709	&	5.1574	&	5.2447	&	5.3258	&	5.3977	&	5.4600	&	5.5133	&	\textbf{5.5587}	&-- \\
			&	4	&	0.0058	&	0.3454	&	0.6696	&	0.9664	&	1.2281	&	1.4517	&	1.6377	&	1.7891	&	\textbf{1.9100}	&	4.6120	&	4.6204	&	4.6358	&	4.6568	&	4.6820	&	4.7097	&	4.7386	&	4.7675	&	\textbf{4.7957}	&-- \\
			&	6	&	3.5939$^{\dagger}$	&	0.1144	&	0.2271	&	0.3362	&	0.4406	&	0.5389	&	0.6304	&	0.7147	&	\textbf{0.7915}	&	4.5290	&	4.5305	&	4.5341	&	4.5396	&	4.5467	&	4.5554	&	4.5652	&	4.5759	&	\textbf{4.5873}	&5\\
			&	8	&	9.3472$^{\dagger}$	&	0.0510	&	0.1015	&0.1513	&	0.2000	&	0.2473	&	0.2928	&	0.3365	&	\textbf{0.3781}	&	4.5083	&	4.5087	&	4.5099	&	4.5118	&	4.5144	&	4.5177	&	4.5215	&	4.5259	&	\textbf{4.5306}	&8\\
			&	20	&	-3.1044$^{\dagger}$	&	0.0036	&	0.0071	&	0.0106	&	0.0142	&	0.0177	&	0.0212	&	0.0247	&	\textbf{0.0282}	&	4.4966	&	4.4966	&	4.4966	&	4.4967	&	4.4968	&	4.4969	&	4.4970	&	4.4972	&	\textbf{4.4974}	&12\\\hline
			8	&	2	&	0.7969	&	2.9783	&	3.9087	&	4.1546	&	\textbf{4.2707}	&	4.0923$^{\circleddash}$	&	3.9039$^{\circleddash}$	&	3.7191$^{\circleddash}$	&	3.5441$^{\circleddash}$	&	5.4132	&	5.5653	&	5.7211	&	5.8451	&	5.9363	&	6.0017	&	6.0483	&	6.0813	&	\textbf{6.1045}	 & -- \\
			&	4	&	0.0080	&	0.6420	&	1.1792	&	1.5782	&	1.8475	&	2.0157	&	2.1115	&	2.1578	&	\textbf{2.1921}	&	4.8188	&	4.8458	&	4.8931	&	4.9495	&	5.0058	&	5.0573	&	5.1019	&	5.1396	&	\textbf{5.1711}& -- \\
			&	6	&	5.9126$^{\dagger}$	&	0.2228	&	0.4324	&	0.6193	&	0.7782	&	0.9084	&	1.0119	&	1.0920	&	\textbf{1.1524}	&	4.6066	&	4.6124	&	4.6254	&	4.6440	&	4.6661	&	4.6897	&	4.7135	&	4.7364	&	\textbf{4.7578}	&5\\
			&	8	&	1.7388$^{\dagger}$	&	0.1006	&	0.1983	&	0.2903	&	0.3749	&	0.4509	&	0.5179	&	0.5762	&	\textbf{0.6261}	&	4.5401	&	4.5418	&	4.5464	&	4.5534	&	4.5625	&	4.5730	&	4.5845	&	4.5963	&	\textbf{4.6083}	&7\\
			&	20	&	-2.3741$^{\dagger}$	&	0.0071	&	0.0142	&	0.0212	&	0.0281	&	0.0350	&	0.0417	&	0.0484	&	\textbf{0.0548}	&	4.4978	&	4.4978	&	4.4979	&	4.4981	&	4.4984	&	4.4989	&	4.4994	&	4.4999	&	\textbf{4.5006}	&11\\\hline
			16	&	2	&	0.7353	&	\textbf{4.4743}$^\varoplus$		&	4.2399$^\varoplus$		&	3.8668$^\varoplus$		&	3.5069	&	3.1937$^{\circleddash}$	&	2.9273$^{\circleddash}$	&	2.6928$^{\circleddash}$	&	2.4877$^{\circleddash}$	&	5.7142	&	5.9808	&	6.1743	&	6.2756	&	6.3253	&	6.3478	&	\textbf{6.3553}	&	6.3545	&	6.3491	& -- \\
			&	4	&	0.0079	&	1.0863	&	1.7281	&	1.9998	&	\textbf{2.0741}	&	2.0403$^{\circleddash}$	&	1.9784$^{\circleddash}$	&	1.9049$^{\circleddash}$	&	1.8266$^{\circleddash}$	&	5.1061	&	5.1771	&	5.2821	&	5.3722	&	5.4377	&	5.4826	&	5.5127	&	5.5327	&	\textbf{5.5456}	& -- \\
			&	6	&	7.2245$^{\dagger}$	&	0.4132	&	0.7464	&	0.9749	&	1.1148	&	1.1921	&	1.2283	&	1.2384	&	\textbf{1.2387}	&	4.7759	&	4.7949	&	4.8344	&	4.8802	&	4.9230	&	4.9595	&	4.9890	&	5.0125	&	\textbf{5.0309}	& 5\\
			&	8	&	2.9594$^{\dagger}$	&	0.1934	&	0.3660	&	0.5064	&	0.6131	&	0.6902	&	0.7438	&	0.7810	&	\textbf{0.8055}	&	4.6278	&	4.6340	&	4.6497	&	4.6711	&	4.6943	&	4.7169	&	4.7376	&	4.7558	&	\textbf{4.7717}& 7	\\
			&	20	&	-2.1084$^{\dagger}$	&	0.0141	&	0.0280	&	0.0416	&	0.0546	&	0.0671	&	0.0788	&	0.0897	&	\textbf{0.0999}	&	4.5022	&	4.5022	&	4.5027	&	4.5035	&	4.5047	&	4.5062	&	4.5078	&	4.5097	&	\textbf{4.5117}	&10\\\hline
		\end{tabular}
	}
\end{sidewaystable}
%
%

Subsequently, \tab\ref{tab:1} summarizes total potential drop ($|\Delta U|$) on the centreline ($P_0$ to $P_4$; \fig\ref{fig:1}) of the device as a function of $K$, ${S_\text{1}}$, and ${S_\text{rh}}$. Maximum potential  ($|\Delta U|_{\text{max}}$) values at each ${S_\text{1}}$ and $K$ for $0\le {S_\text{rh}}\le 2$ are highlighted as bold data. 
The $|\Delta U|$ decreases with increasing $K$, and the most significant impact of $K$ on $|\Delta U|$ is observed at lowest ${S_\text{1}}=4$ and ${S_\text{rh}}=0$ (\tab\ref{tab:1}).  For instance, as $K$ varied from 2 to 20, $|\Delta U|$ for (${S_\text{rh}}=0$, 1, 2) drops by 99.7\% (from $47.0420$ to $0.1413$), 99.66\% (from $62.5850$ to $0.2118$), 99.56\% (from $64.4670$ to $0.2812$) at $ {S_\text{1}}=4$ and 98.79\% (from $44.9890$ to $0.5441$), 98.45\% (from $52.6360$ to $0.8158$), 97.83\% (from $477470$ to $1.0382$)  at ${S_\text{1}}=16$ (refer \tab\ref{tab:1}).
Further, the variation in $|\Delta U|$ with ${S_\text{1}}$ is maximum at ${S_\text{rh}}=1$ and $K=20$ (\tab\ref{tab:1}). For instance, as $K$ varied from 2 to 20, $|\Delta U|$ for (${S_\text{rh}}=0$, 1, 2) changes by $-4.36\%$ (from $47.0420$ to $44.9890$), $-15.90\%$ (from $62.5850$ to $52.6360$), $-25.94\%$ (from $64.4670$ to $47.7470$) at ${S_\text{1}}=4$ and 285.18\% (from $0.1413$ to $0.5441$), 285.24\% (from $0.2118$ to $0.8158$), 269.16\% (from $0.2812$ to $1.0382$) at ${S_\text{1}}=16$ (refer \tab\ref{tab:1}).

\begin{figure}[!b]
	\centering\includegraphics[width=1\linewidth]{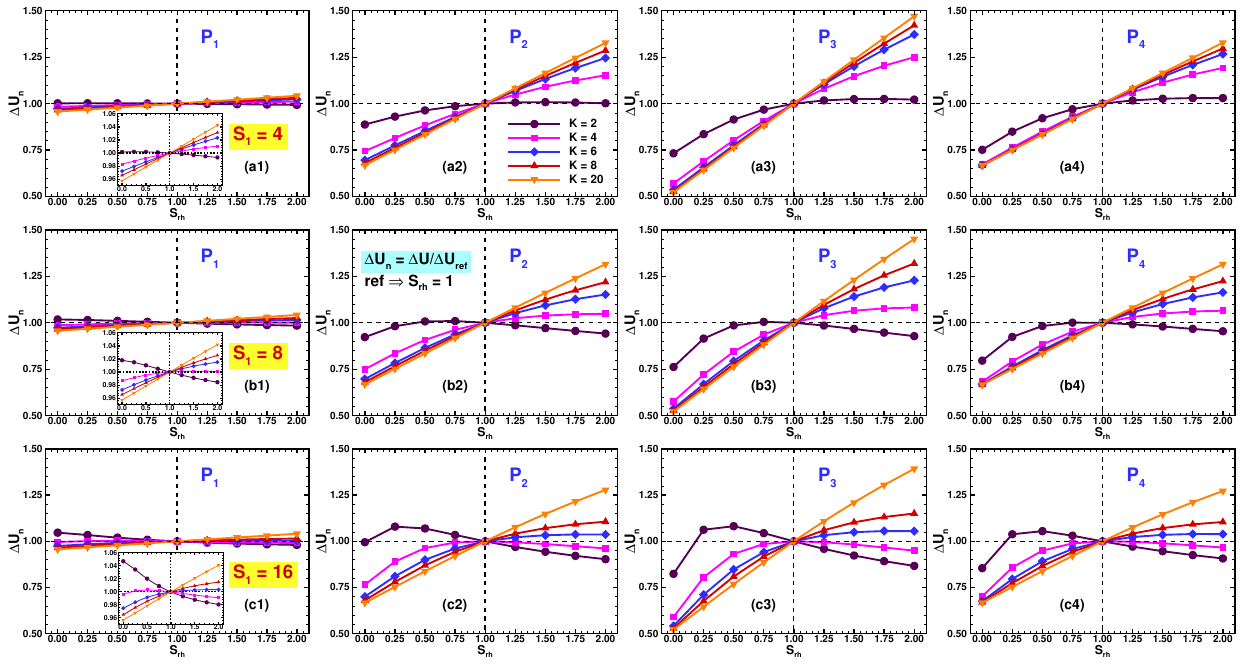}
	\caption{Normalized potential drop ($\Delta U_\text{n}$, \eqn\ref{eq:11a}) variation with ${S_\text{rh}}$ on centreline locations (P$_{\text{j}}$; \fig\ref{fig:1}) of heterogeneously charged microfluidic device for $2\le K\le 20$ and $4\le  {S_\text{1}}\le 16$.}
	\label{fig:4}
\end{figure} 
%
Furthermore, the effect of surface charge heterogeneity (${S_\text{rh}}$) on $|\Delta U|$ is maximum at weak EV conditions (i.e., $ {S_\text{1}}=4$, $K=20$). For instance, $|\Delta U|$ reduces for ($ {S_\text{1}}=4$, 8, 16) by 24.84\% (from $62.5850$ to $47.0420$), 20.37\% (from $64.0740$ to $51.0220$), 14.53\% (from $52.6360$ to $44.9890$) at $K=2$, and 33.29\% (from $0.2118$ to $0.1413$), 33.30\% (from $0.4202$ to $0.2803$), 33.30\% (from $0.8158$ to $0.5441$) at $K=20$ with decreasing ${S_\text{rh}}<1$ (from 1 to 0); on the other hand, with increasing $S_\text{rh}>1$ (from 1 to 2), $|\Delta U|$ changes by 3.01\% (from $62.5850$ to $64.4670$), -4.56\% (from $64.0740$ to $61.1550$), -9.29\% (from $52.6360$ to $47.7470$) at $K=2$, and 32.81\% (from $0.2118$ to $0.2812$), 31.45\% (from $0.4202$ to $0.5524$), 27.26\% (from $0.8158$ to $1.0382$) at $K=20$. The overall influence of charge-heterogeneity ($0\le{S_\text{rh}}\le 2$) on the values of $|\Delta U|$ is noted for (${S_\text{1}}=4$, 8, 16) as 37.04\% (from $47.0420$ to $64.4670$), 19.86\% (from $51.0220$ to $61.1550$), 6.13\% (from $44.9890$ to $47.7470$) at $K=2$, and 99.09\% (from $0.1413$ to $0.2812$), 97.07\% (from $0.2803$ to $0.5524$), 90.81\% (from $0.5441$ to $1.0382$) at $K=20$ (refer \tab\ref{tab:1}). In general, enhancement in $|\Delta U|$ is noted with increasing of both ${S_\text{1}}$ and ${S_\text{rh}}$, however, $|\Delta U|$ decreases at higher ${S_\text{1}}$ and ${S_\text{rh}}$. It is due to significantly stronger charge attractive force near the device walls at higher ${S_\text{rh}}$ and ${S_\text{1}}$ resists the convective flow of ions, thus, reduces streaming potential and hence $|\Delta U|$ (refer \fig\ref{fig:3} and \tab\ref{tab:1}).
%

Subsequently, the relative impact of surface charge heterogeneity on the electrical potential drop is analyzed in \fig\ref{fig:4}, which shows the normalized potential drop ($\Delta U_\text{n}$, \eqn\ref{eq:11a}) variation with ${S_\text{rh}}$ on the centreline points (P$_{\text{j}}$, where $1\le \text{j}\le 4$; \fig\ref{fig:1}) of considered microfluidic device for the considered ranges of conditions (\tab\ref{tab:1b}). The normalized values show a complex dependency on the flow governing parameters. For instance, $\Delta U_\text{n}$ increases with decreasing $K$ (EDL thickening) for ${S_\text{rh}}<1$ followed by reverse trends for ${S_\text{rh}}>1$ at all points P$_{\text{j}}$ (\fig\ref{fig:4}).  Further, the maximum variation in $\Delta U_\text{n}$ with $K$ is obtained at highest ${S_\text{1}}$ and $ {S_\text{rh}}$ at P$_3$. For instance, $\Delta U_\text{n}$ enhances maximally for (P$_1$, P$_2$, P$_3$, P$_4$) by 6.16\% (from $0.9806$ to $1.0410$), 41.60\% (from $0.9031$ to $1.2788$), 60.72\% (from $0.8667$ to $1.3930$), 40.29\% (from $0.9071$ to $1.2726$), respectively with increasing $K$ from 2 to 20 at $ {S_\text{1}}=16$, and ${S_\text{rh}}=2$ (refer \fig\ref{fig:4}c). Similarly, $\Delta U_\text{n}$ increases with increasing ${S_\text{1}}$ for ${S_\text{rh}}<1$, but it decreases with increment of $ {S_\text{1}}$ for ${S_\text{rh}}>1$ for all centreline points of device; the relative impact of ${S_\text{1}}$ on $\Delta U_\text{n}$ is maximum at lower $K$ and ${S_\text{rh}}$ at P$_3$. For instance, $\Delta U_\text{n}$ increases maximally by 3.23\% (from $1.0020$ to $1.0344$), 16.06\% ($0.9302$ to $1.0796$), 27.17\% (from $0.8358$ to $1.0628$), 22.31\% (from $0.8490$ to $1.0384$) at P$_1$, P$_2$, P$_3$, P$_4$, respectively when $ {S_\text{1}}$ changes from 4 to 16 at $K=2$, ${S_\text{rh}}=0.25$ (refer \fig\ref{fig:4}).

Further, $\Delta U_\text{n}$ enhances with increasing ${S_\text{rh}}$, but reverse trends are observed at higher ${S_\text{rh}}$ and lower $K$ (\fig\ref{fig:4}). Because EDLs overlap at higher ${S_\text{rh}}$ and lower $K$, the  {advection} of excess ions in the fluid is impeded. The relative effect of ${S_\text{rh}}$ on $\Delta U_\text{n}$ is observed maximum at highest $K$ and lowest ${S_\text{1}}$ at P$_3$ (\fig\ref{fig:4}). For instance, maximum increment in the values of $\Delta U_\text{n}$ is noted for P$_1$, P$_2$, P$_3$, P$_4$ as 8.88\% (from $0.9574$ to $1.0424$), 98.63\% (from $0.6686$ to $1.3274$), 181.24\% (from $0.5228$ to $1.4704$), 99.09\% (from $0.6671$ to $1.3281$), respectively when ${S_\text{rh}}$ varies from 0 to 2 at weak EV ($K=20$, ${S_\text{1}}=4$) condition (refer \fig\ref{fig:4}a). Thus, it is noted that maximum variation in $\Delta U_\text{n}$ with dimensionless parameters ($K$, $ {S_\text{1}}$, ${S_\text{rh}}$) is obtained at P$_3$ than other centreline locations (P$_1$, P$_2$, P$_4$) of the microfluidic device because P$_3$ attributes the non-equilibrium point, i.e., relative variation of surface charge density ($S_\text{k}$) is higher in heterogeneous (${S_\text{2}}$, \eqn\ref{eq:4}) than downstream (${S_\text{1}}$) section, for given ranges of condition (\fig\ref{fig:4}). 
Furthermore, the Poisson's equation (\eqn\ref{eq:1}) relates the total potential ($U$) distribution with excess charge ($n^\ast$), the subsequent section analyzes the excess charge ($n^\ast$) distribution for wide ranges of dimensionless parameters ($K$, $ {S_\text{1}}$, $ {S_\text{rh}}$).  
%
\subsection{Excess charge ($n^\ast$) distribution}
\label{sec:charge}
\begin{figure}[!b]
	\centering\includegraphics[width=1\linewidth]{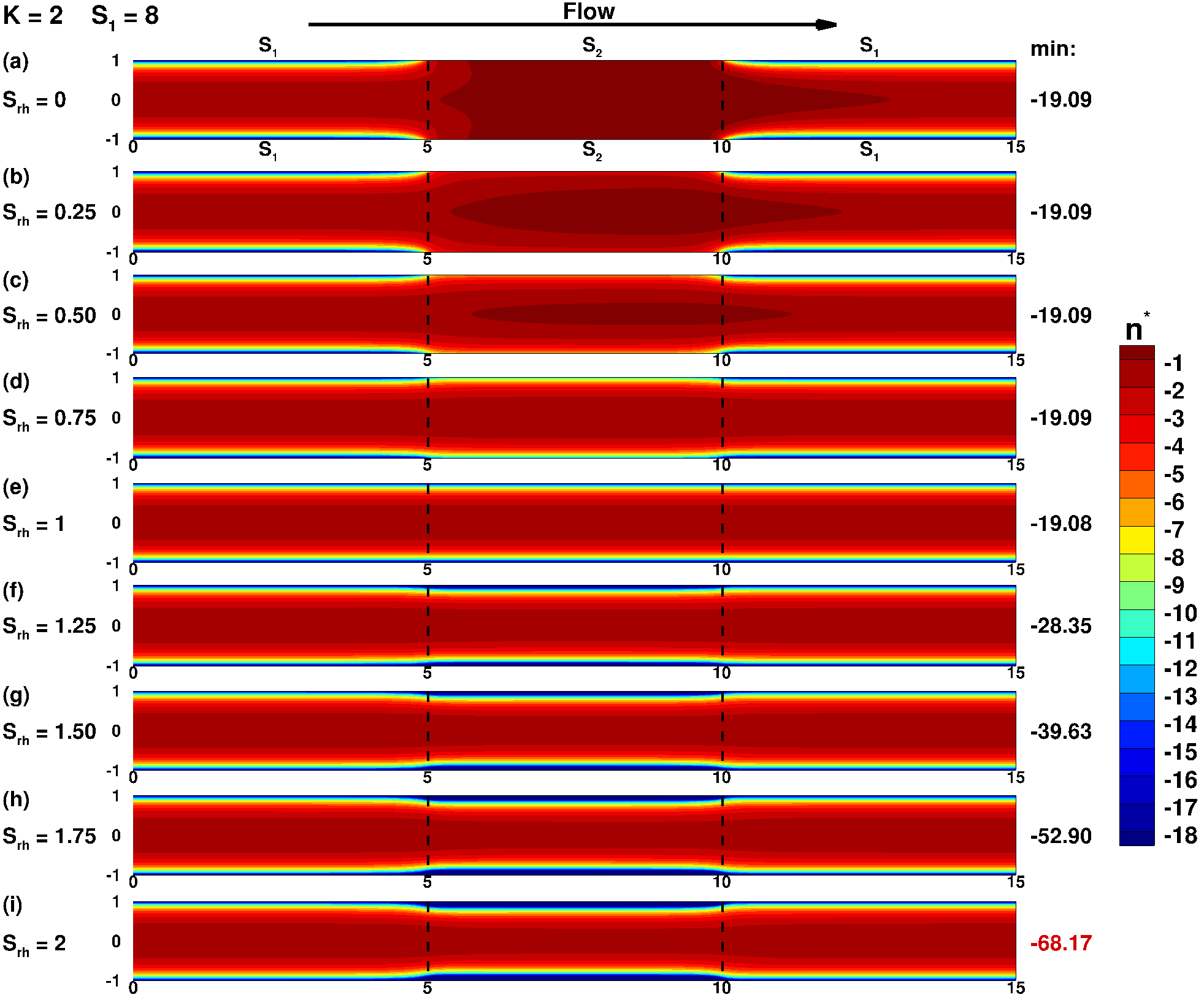}
	\caption{Influence of charge heterogeneity ($0\le  {S_\text{rh}}\le 2$) on the excess charge ($n^\ast$) distribution for the fixed condition ($S_\text{1}=8$ and $K=2$).}
	\label{fig:5}
\end{figure}

%
\noindent 
An influence of charge heterogeneity ($0\le  {S_\text{rh}}\le 2$) on the excess charge ($n^\ast$, \eqn\ref{eq:1}) distribution in the heterogeneously charged microfluidic device for the fixed conditions (${S_\text{1}}=8$, $K=2$) is displayed in \fig\ref{fig:5}; qualitatively similar profiles observed at the other conditions (\tab\ref{tab:1b}) are not shown here. 
Excess charge is seen as negative ($n^\ast<0$), except for few conditions (lower ${S_\text{rh}}$ and highest $K=20$), throughout the channel for given ranges of conditions due to the positively charged surfaces of the device (\fig\ref{fig:5}). 
The dense clustering of $n^\ast$ is obtained near the walls of homogeneously charged ($ {S_\text{rh}}=1$) microfluidic device (\fig\ref{fig:5}e). However, clustering of $n^\ast$ is further enhanced in the heterogeneous section of device for ${S_\text{rh}}>1$ followed by reverse trends for ${S_\text{rh}}<1$ as compared to ${S_\text{rh}}=1$ (\fig\ref{fig:5}). Thus, the heterogeneous section (between lines \textit{a} and \textit{b}, \fig\ref{fig:1}) of the device behaves like a sudden contraction and expansion for ${S_\text{rh}}>1$ and ${S_\text{rh}}<1$, respectively, than upstream/downstream section. It is because of variation in the charge attractive force near the walls of the heterogeneous section with varying ${S_\text{rh}}$. Further, $n^\ast$ decreases with increasing ${S_\text{rh}}$ due to strengthening in the electrostatic force close to the device walls, which enhances the excess charge distribution in the device (\fig\ref{fig:5}). Minimum value of $n^\ast$ is observed as $-68.17$ (at ${S_\text{rh}}=2$) for $K=2$ and ${S_\text{1}}=8$ (\fig\ref{fig:5}i). However, overall minimum value of $n^\ast$ is noted as $-260.9$ (at ${S_\text{rh}}=2$, $K=2$, ${S_\text{1}}=16$). 
\begin{figure}[!b]
	\centering\includegraphics[width=1\linewidth]{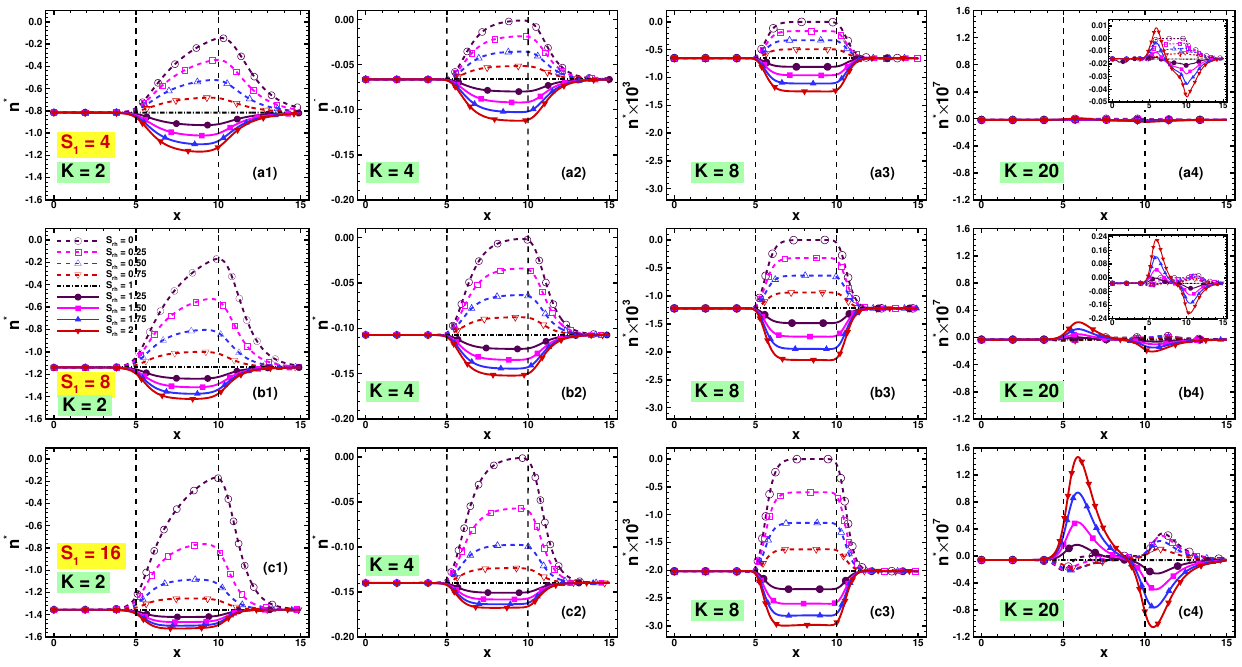}
	\caption{Excess charge ($n^\ast$, \eqn\ref{eq:1}) variation on the centreline (P$_0$ to P$_4$; \fig\ref{fig:1}) of heterogeneously charged microfluidic device for dimensionless parameters (\tab\ref{tab:1b}).}
	\label{fig:6}
\end{figure} 
%
Furthermore, \fig\ref{fig:6} depicts excess charge ($n^\ast$, \eqn\ref{eq:1}) variation on the centreline (P$_0$ to P$_4$; \fig\ref{fig:1}) of the microfluidic device for governing parameters ($K$, ${S_\text{1}}$, ${S_\text{rh}}$; \tab\ref{tab:1b}). At ${S_\text{rh}}=1$, excess charge is constant along the length of the device for fixed $K$ and $ {S_\text{1}}$. Qualitative variation of $n^\ast$ on the centreline (P$_0$ to P$_4$) of the device depicts similar trends with the literature \citep{davidson2007electroviscous,dhakar2022electroviscous, dhakar2023cfd,dhakar2024influence} for ${S_\text{rh}}>1$ and opposite trends for ${S_\text{rh}}<1$, respectively for given ranges of $K$ and $ {S_\text{1}}$. The $n^\ast$ is maximum for $ {S_\text{rh}}<1$ and minimum for $ {S_\text{rh}}>1$ in the heterogeneous than other sections of device (\fig\ref{fig:6}). It is due to charge attractive force variation close to the walls of the heterogeneous section with $ {S_\text{rh}}$. The critical value of excess charge ($n^\ast_{\text{c}}$) is expressed as the minimum (or maximum) value of $n^\ast$ for given ranges of conditions. The $n^\ast_{\text{c}}$ decreases with decreasing $K$ (EDL thickening). Further, $n^\ast_\text{c}$ decreases with increasing both $S_\text{1}$ and ${S_\text{rh}}$ (\fig\ref{fig:6}). Maximum variation in $n^\ast_{\text{c}}$ is observed  from $-1.5294\times10^{-8}$ to $-3\times10^{-3}$ when ${S_\text{rh}}$ varies from 0 to 2 at $K=8$ and ${S_\text{1}}=16$ (\fig\ref{fig:6}c3).

Subsequently, \tab\ref{tab:1} summarizes the critical value of excess charge ($n^\ast_{\text{c}}$) on the centreline (P$_0$ to P$_4$; \fig\ref{fig:1}) of the device as a function of $K$, ${S_\text{1}}$, and ${S_\text{rh}}$. The critical values are either maximum (noted with superscript $\varoplus$) or minimum (noted with superscript $\circleddash$). 
The smallest value of $n^\ast_{\text{c}}$ for $0\le  {S_\text{rh}}\le 2$ at each fixed ${S_\text{1}}$ and $K$ are highlighted with bold data. 
 The $n^\ast_{\text{c}}$ increases with increasing $K$ (i.e., EDL thinning) and approaches zero {($n^\ast_{\text{c}}\rightarrow 0$)} for $K\ge 8$ over the ranges of $ {S_\text{1}}$ and $ {S_\text{rh}}$ (\tab\ref{tab:1}). Due to charge heterogeneity (${S_\text{rh}}$), few positive values are seen at higher $K$, higher ${S_\text{1}}$, and small ${S_\text{rh}}$. The change in the value of $n^\ast_{\text{c}}$ with $K$ is maximum at lowest ${S_\text{rh}}=0$ and ${S_\text{1}}=4$. For instance, $n^\ast_{\text{c}}$ reduces  (from $-0.1453$ to $-6.9646\times10^{-9}$), (from $-0.8159$ to $-0.0007$), (from $-1.1663$ to $-0.0013$) for ${S_\text{1}}=4$ and (from $-0.1684$ to $-1.5294\times10^{-8}$), (from $-1.3549$ to $-0.0020$), (from $-1.5241$ to $-0.0030$) for ${S_\text{1}}=16$ at (${S_\text{rh}}=0$, 1, 2) when $K$ varies from 2 to 8. The variation in $n^\ast_\text{c}$ with ${S_\text{1}}$ is maximum at ${S_\text{rh}}=1$ and higher $K=8$ (refer \tab\ref{tab:1}). For instance, increment in the values of $n^\ast_{\text{c}}$ are noted for (${S_\text{rh}}=0$, 1, 2) as 15.91\% (from $-0.1453$ to $-0.1684$), 66.07\% (from $-0.8159$ to $-1.3549$), 30.68\% (from $-1.1663$ to $-1.5241$) at $K=2$ and (from $-6.9646\times10^{-9}$ to $-1.5294\times10^{-8}$), (from $-0.0007$ to $-0.0020$), (from $-0.0013$ to $-0.0030$) at $K=8$ when ${S_\text{1}}$ varies from $4$ to $16$ (refer \tab\ref{tab:1}). 

Further, the effect of $S_\text{rh}$ on $n^\ast_{\text{c}}$ is obtained maximum at higher ${S_\text{1}}$ and $K$ (\tab\ref{tab:1}). For instance, $n^\ast_{\text{c}}$ for ${S_\text{1}}=4$, 8, 16 increases by 82.19\% (from $-0.8159$ to $-0.1453$), 85.18\% (from $-1.1391$ to $-0.1688$), 87.57\% (from $-1.3549$ to $-0.1684$) at $K=2$ and (from $-0.0007$ to $-6.9646\times10^{-9}$), (from $-0.0012$ to $-1.2272\times10^{-8}$), (from -0.0020 to $-1.5294\times10^{-8}$) at $K=8$ with decreasing $S_\text{rh}$ from 1 to 0; on the other hand, corresponding decrement in $n^\ast_{\text{c}}$ is obtained with increasing $S_\text{rh}$ from 1 to 2 as 42.96\% (from $-0.8159$ to $-1.1663$), 24.74\% (from $-1.1391$ to $-1.4209$), 12.49\% (from $-1.3549$ to $-1.5241$) at $K=2$ and 91.97\% (from $-0.0007$ to $-0.0013$), 75.73\% (from $-0.0012$ to $-0.0021$), 48.46\% (from $-0.0020$ to $-0.0030$) at $K=8$. With overall increasing charge-heterogeneity (${S_\text{rh}}$) from 0 to 2,  $n^\ast_{\text{c}}$ enhances significantly (from $-0.1453$ to $-1.1663$), (from $-0.1688$ to $-1.4209$), (from $-0.1684$ to $-1.5241$) at $K=2$ and enormously (from $-6.9646\times10^{-9}$ to $-0.0013$), (from $-1.2272\times10^{-8}$ to $-0.0021$), (from $-1.5294\times10^{-8}$ to $-0.0030$) at $K=8$ for (${S_\text{1}}=4$, 8, 16) (refer \tab\ref{tab:1}). In general, decrement in $n^\ast_\text{c}$ is observed with enhancing both ${S_\text{rh}}$ and $ {S_\text{1}}$ due to strengthening in the charge attractive force in the close vicinity of microfluidic device walls with increasing ${S_\text{rh}}$ and ${S_\text{1}}$ (\fig\ref{fig:6} and \tab\ref{tab:1}).  
\begin{figure}[!b]
	\centering\includegraphics[width=1\linewidth]{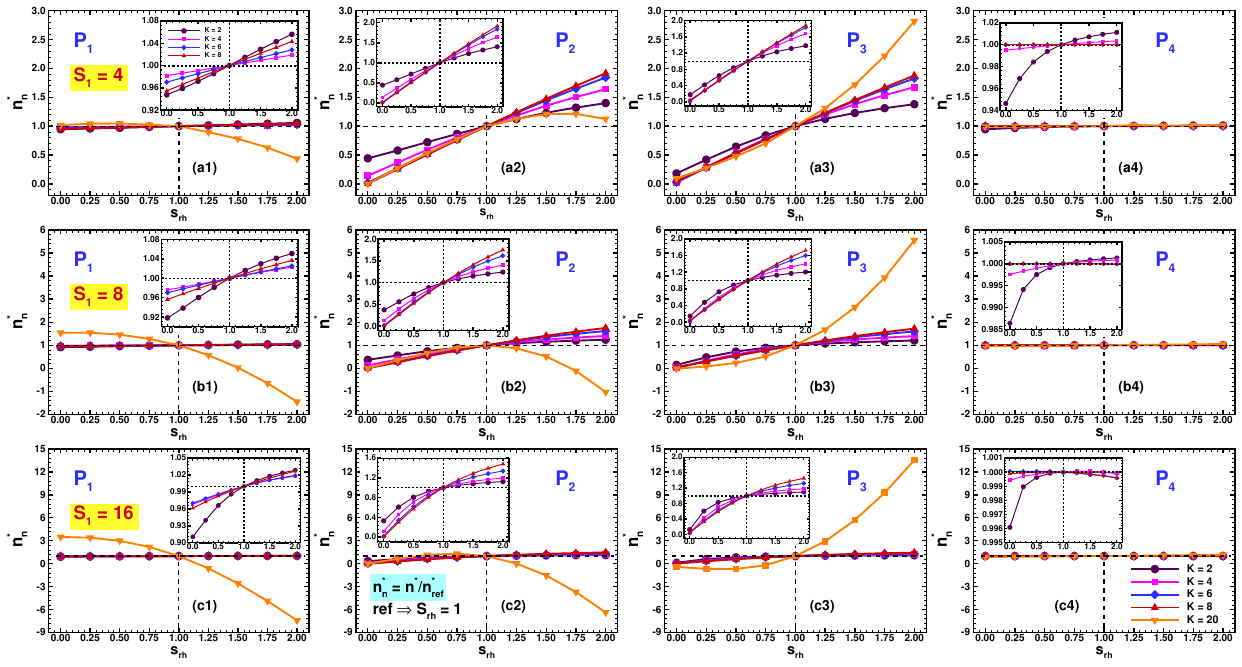}
	\caption{Normalized excess charge ($n^\ast_\text{n}$, \eqn\ref{eq:11a}) variation on centreline points (P$_\text{j}$; \fig\ref{fig:1}) of heterogeneously charged microfluidic device with ${S_\text{rh}}$, $K$ and ${S_\text{1}}$.}
	\label{fig:7}
\end{figure} 
%

Furthermore, \fig\ref{fig:7} depicts normalized excess charge ($n^\ast_{\text{n}}$, \eqn\ref{eq:11a}) variation with ${S_\text{rh}}$ on centreline locations (P$_{\text{j}}$; refer \fig\ref{fig:1}) of a device for considered parameters (\tab\ref{tab:1b}). 
The $n^\ast_{\text{n}}$ values show complex dependency on governing parameters, i.e., it increases with decreasing $K$ or EDL thickening for ${S_\text{rh}}<1$ followed by opposite trends for ${S_\text{rh}}>1$ for  at the centreline points of the device (\fig\ref{fig:7}). The change in $n^\ast_{\text{n}}$ with $K$ is obtained maximum at highest ${S_\text{rh}}$ and ${S_\text{1}}$ at P$_3$ (\fig\ref{fig:7}). For instance, $n^\ast_{\text{n}}$ varies (from $1.0283$ to $-7.4226$), (from $1.1248$ to $-6.3695$), (from $1.1013$ to $13.5428$), (from $0.9999$ to $1.1665$) on points (P$_1$, P$_2$, P$_3$, P$_4$), respectively when $K$ varies from 2 to 20 at ${S_\text{1}}=16$ and ${S_\text{rh}}=2$ (refer \fig\ref{fig:7}c). The $n^\ast_{\text{n}}$ decreases with increasing ${S_\text{1}}$ for all points; maximum variation in $n^\ast_{\text{n}}$ with ${S_\text{1}}$ is observed at highest $K$ and ${S_\text{rh}}$ at P$_2$ (\fig\ref{fig:7}). For instance, change in the values of $n^\ast_{\text{n}}$ are recorded at points (P$_1$, P$_2$, P$_3$, P$_4$) as (from $0.4365$ to $-7.4226$), (from $1.1252$ to $-6.3695$), (from $2.8171$ to $13.5428$), (from $1.0147$ to $1.1665$) at $K=20$ and ${S_\text{rh}}=2$, respectively when ${S_\text{1}}$ varies from 4 to 16 (refer \fig\ref{fig:7}). 

Further, $n^\ast_{\text{n}}$ enhances with increasing $ {S_\text{rh}}$, but it attributes opposite trends at lower $K$ (thick EDL) and higher $ {S_\text{rh}}$ (\fig\ref{fig:7}). It is because EDL occupy the greater fraction of microchannel at higher ${S_\text{rh}}$ and lower $K$, reducing the effective excess charge in the device. The relative impact of ${S_\text{rh}}$ on $n^\ast_{\text{n}}$ is observed maximum at highest ${S_\text{1}}$ and $K$ at P$_2$ (\fig\ref{fig:7}). For instance, variation in the values of $n^\ast_{\text{n}}$ is noted  (from $3.4936$ to $-7.4226$), (from $0.0030$ to $-6.3695$), (from $-0.4151$ to $13.5428$), (from $0.8907$ to $1.1665$) for points (P$_1$, P$_2$, P$_3$, P$_4$), respectively when ${S_\text{rh}}$ changes from 0 to 2 at $K=20$ and ${S_\text{1}}=16$ (refer \fig\ref{fig:7}c). Thus, it is observed that maximum variation in $n^\ast_{\text{n}}$ with governing parameters ($K$, ${S_\text{1}}$, ${S_\text{rh}}$) is recorded at points P$_2$ and P$_3$ compared with other locations (P$_1$, P$_4$). It is because the potential has shown significant variation on P$_3$ than other points (refer \fig\ref{fig:4}). Thus, from \eqn(\ref{eq:1}), variation in $n^\ast$ and hence $n^\ast_\text{n}$ are greater at P$_2$ and P$_3$ than other centreline points (\fig\ref{fig:7}). At $K=20$, drastic changes are obtained in $n^\ast_{\text{n}}$ with ${S_\text{rh}}$ at mainly P$_1$, P$_2$, and P$_3$ for $4\le {S_\text{1}}\le 16$ (\fig\ref{fig:7}).
%
\subsection{Induced electric field strength ($E_{\text{x}}$)}
\label{sec:electric}
\begin{figure}[!b]
	\centering\includegraphics[width=1\linewidth]{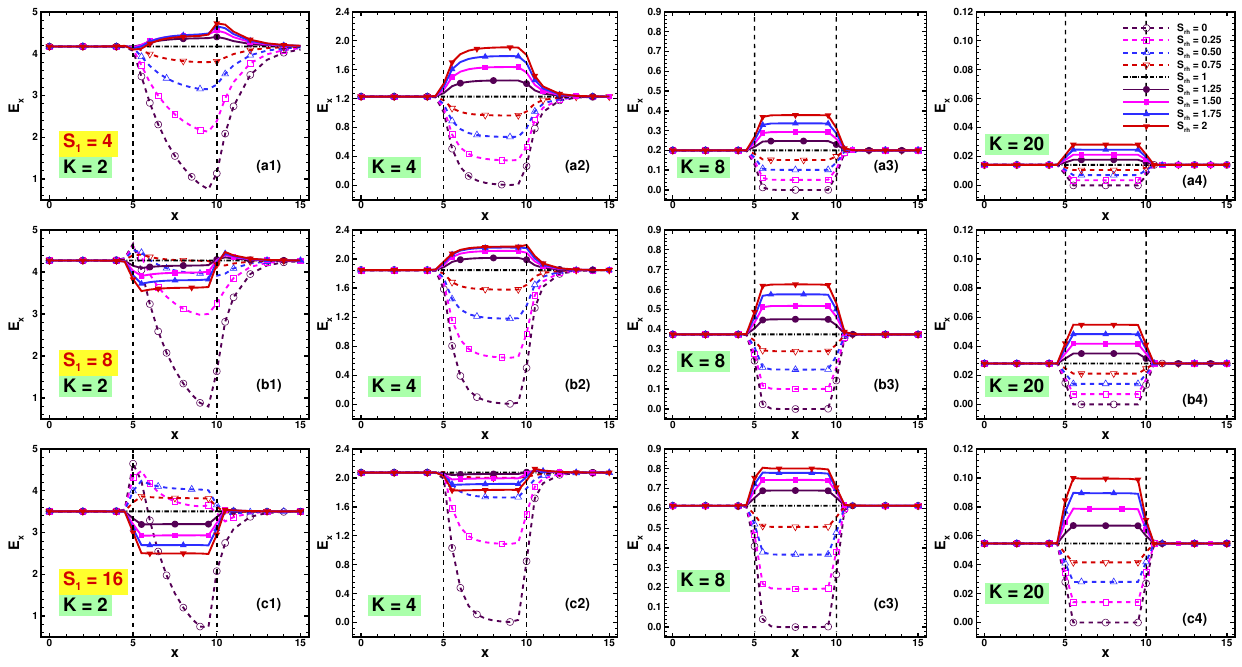}
	\caption{Induced electric field strength ($E_\text{x}$) variation on the centreline ($P_0$ to $P_4$; \fig\ref{fig:1}) of heterogeneously charged microfluidic device for dimensionless parameters ($K$, $ {S_\text{1}}$, $ {S_\text{rh}}$; \tab\ref{tab:1b}).}
	\label{fig:8}
\end{figure}

%
The {advection} of excess charge ($n^\ast$), by an imposed pressure-driven flow (PDF) in the microfluidic device, develops an induced electric field strength ($E_\text{x}=-\partial_x U$, \eqn\ref{eq:2}) in the axial flow direction. \fig\ref{fig:8} shows induced electric field strength ($E_\text{x}$) variation on the centreline (P$_0$ to P$_4$; \fig\ref{fig:1}) of heterogeneously charged device for the considered range of the flow governing parameters ($K$, ${S_\text{1}}$, ${S_\text{rh}}$; \tab\ref{tab:1b}).  {The field strength} is uniform throughout the homogeneously charged (${S_\text{rh}}=1$) microfluidic device, irrespective of $K$ and ${S_\text{1}}$ (\fig\ref{fig:8}). The $E_\text{x}$ depicts similar qualitative trends with the literature \citep{dhakar2022electroviscous,dhakar2023cfd,dhakar2024influence} for ${S_\text{rh}}>1$ and opposite for ${S_\text{rh}}<1$, respectively, except at lower $K$ and higher ${S_\text{1}}$. For instance, $E_\text{x}$ is maximum for ${S_\text{rh}}>1$ and minimum for ${S_\text{rh}}<1$ in the heterogeneous than other sections excluding at lower $K$ and higher ${S_\text{1}}$ condition (\fig\ref{fig:8}). It is due to {complex variations} in the electrostatic force close to the heterogeneous region of the walls with varying ${S_\text{rh}}$. The critical value of induced electric field strength ($E_\text{x,c}$) is  {recorded} as either the maximum or minimum value of $E_\text{x}$ for given  condition. The $E_\text{x,c}$ increases with decreasing $K$ or EDL thickening (\fig\ref{fig:8}). Further, $E_\text{x,c}$ enhances with increasing ${S_\text{1}}$ and ${S_\text{rh}}$; but at higher ${S_\text{1}}$ and ${S_\text{rh}}$, it has shown opposite trends with increasing ${S_\text{1}}$ and ${S_\text{rh}}$. Maximum variation in the value of $E_\text{x,c}$ is observed from $-3.1044\times10^{-12}$ to $2.82\times10^{-2}$ when ${S_\text{rh}}$ varies from 0 to 2 at weak EVF ($K=20$, ${S_\text{1}}=4$) condition (refer \fig\ref{fig:8}a4).
%

Subsequently, \tab\ref{tab:1} summarizes critical values of induced electric field strength ($E_\text{x,c}$) on the centreline (P$_0$ to P$_4$; \fig\ref{fig:1}) of the device as a function of $K$, ${S_\text{1}}$, and ${S_\text{rh}}$. The critical values are either maximum (noted with superscript $\varoplus$) or minimum (noted with superscript $\circleddash$). The lowest  $E_\text{x,c}$ for $0\le  {S_\text{rh}}\le 2$ at each $ {S_\text{1}}$ and $K$ {is} highlighted with bold data.  The $E_\text{x,c}$ decreases with increasing $K$; maximum variation in $E_\text{x,c}$ with $K$ is obtained at lowest ${S_\text{1}}=4$ (\tab\ref{tab:1}). For instance, $E_\text{x,c}$ decreases  (from $0.7542$ to $-3.1044\times10^{-12}$), (from $4.1720$ to $0.0142$), (from $4.7252$ to $0.0282$) at ${S_\text{1}}=4$ and (from $0.7353$ to $-2.1084\times10^{-10}$), (from $3.5069$ to $0.0546$), (from $2.4877$ to $0.0999$) at ${S_\text{1}}=16$ for (${S_\text{rh}}=0$, 1, 2) with increasing $K$ from 2 to 20 at ${S_\text{1}}=4$ and 16 (refer \tab\ref{tab:1}). The maximum changes in $E_\text{x,c}$ with ${S_\text{1}}$ are noted at highest $K=20$ and lowest ${S_\text{rh}}=0$ (\tab\ref{tab:1}). For instance, $E_\text{x,c}$ varies by $-2.50\%$ (from $0.7542$ to $0.7353$), $-15.94\%$ (from $4.1720$ to $3.5069$), $-47.35\%$ (from $4.7252$ to $2.4877$) at $K=2$ and (from $-3.1044\times10^{-12}$ to $-2.1084\times10^{-10}$), (from $0.0142$ to $0.0546$), (from $0.0282$ to $0.0999$) at $K=20$ for (${S_\text{rh}}=0$, 1, 2)  when ${S_\text{1}}$ varies from 4 to 16 (refer \tab\ref{tab:1}). The impact of ${S_\text{rh}}$ on $E_\text{x,c}$ is maximum at weak EVF (${S_\text{1}}=4$, $K=20$) condition (\tab\ref{tab:1}). For instance, $E_\text{x,c}$ reduces for (${S_\text{1}}=4$, 8, 16) by 81.92\%, 81.34\%, 79.03\% at $K=2$ and (from $0.0142$ to $-3.1044\times10^{-12}$), (from $0.0281$ to $-2.3741\times10^{-11}$), (from $0.0546$ to $-2.1084\times10^{-10}$) at $K=20$ with decreasing ${S_\text{rh}}$ from 1 to 0; on the other hand, $E_\text{x,c}$ varies with increasing ${S_\text{rh}}$ from $1$ to $2$ by 13.26\%, $-17.01\%$, $-29.06\%$ at $K=2$ and 98.62\%, 94.81\%, 82.82\% at $K=20$. Overall variation in the values of $E_\text{x,c}$ is recorded for (${S_\text{1}}=4$, 8, and 16) as 526.56\% (from 0.7542 to 4.7252), 344.75\% (from 0.7969 to 3.5441), 238.33\% (from 0.7353 to 2.4877) at $K=2$ and (from $-3.1044\times10^{-12}$ to 0.0282), (from $-2.3741\times10^{-11}$ to 0.0548), (from $-2.1084\times10^{-10}$ to 0.0999) at $K=20$ with increasing ${S_\text{rh}}$ from $0$ to $2$ (refer \tab\ref{tab:1}).

Broadly, $E_\text{x,c}$ has depicted complex dependency on the surface charge density (${S_\text{1}}$ and $ {S_\text{rh}}$). Increment in the values of $E_\text{x,c}$ is obtained with increasing both ${S_\text{rh}}$ and ${S_\text{1}}$, but it has shown decrement in $E_\text{x,c}$ with increasing ${S_\text{1}}$ and ${S_\text{rh}}$ at higher ${S_\text{rh}}$, higher $ {S_\text{1}}$ and lower $K$. It is because strengthening in the electrostatic force increases the available $n^\ast$ (refer \fig\ref{fig:6}) for transport in the channel, which enhances $E_\text{x,c}$ with increasing ${S_\text{1}}$ and ${S_\text{rh}}$. However, remarkably stronger charge attractive force and thick EDL are obtained at higher ${S_\text{1}}$, ${S_\text{rh}}$ and lower $K$, thus, it resists the excess ions flow in the device and decreases $E_\text{x,c}$ (refer \fig\ref{fig:8} and \tab\ref{tab:1}). 
\begin{figure}[!b]
	\centering\includegraphics[width=1\linewidth]{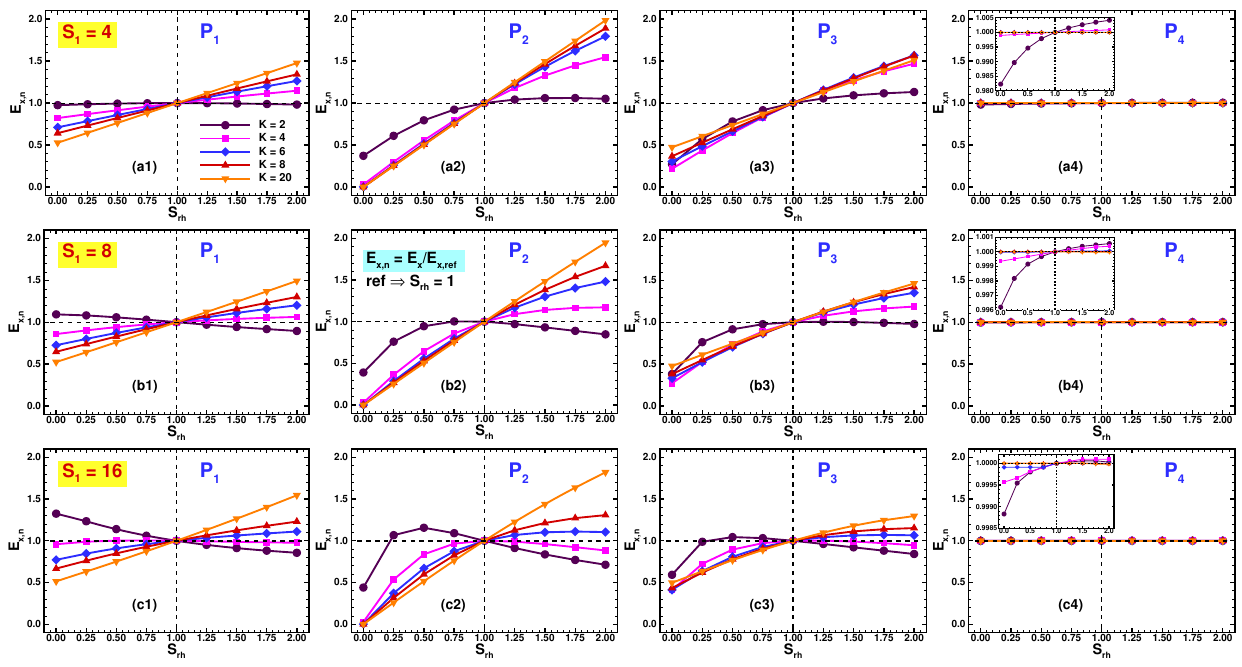}
	\caption{Normalized induced electric field strength ($E_\text{x,n}$, \eqn\ref{eq:11a}) variation with ${S_\text{rh}}$ on the centreline locations (P$_\text{j}$; \fig\ref{fig:1}) of heterogeneously charged microfluidic device for $2\le K\le 20$ and $4\le  {S_\text{1}}\le 16$.}
	\label{fig:9}
\end{figure} 
%

\fig\ref{fig:9} depicts normalized induced electric field strength ($E_\text{x,n}$, \eqn\ref{eq:11a}) variation with ${S_\text{rh}}$ on centreline locations (P$_{\text{j}}$; \fig\ref{fig:1}) of the device for $2\le K\le 20$ and $4\le {S_\text{1}}\le 16$. The $E_\text{x,n}$ shows complex dependency on governing parameters ($K$, ${S_\text{1}}$, ${S_\text{rh}}$) at the selected locations. The $E_\text{x,n}$ increases with decreasing $K$ for ${S_\text{rh}}<1$ followed by opposite trends for ${S_\text{rh}}>1$ for all centreline points (\fig\ref{fig:9}). Maximum increment in $E_\text{x,n}$ is noted at highest ${S_\text{rh}}$ and ${S_\text{1}}$ at P$_2$ (\fig\ref{fig:9}). For instance, $E_\text{x,n}$ enhances by $80.59\%$, $155.42\%$, $54.07\%$, 0\% for (P$_1$, P$_2$, P$_3$, P$_4$), respectively when $K$ varies from 2 to 20 at ${S_\text{1}}=16$ and ${S_\text{rh}}=2$ (refer \fig\ref{fig:9}c). The $E_\text{x,n}$ increases with increasing ${S_\text{1}}$ for ${S_\text{rh}}<1$, but it decreases for ${S_\text{rh}}>1$ for all centreline locations; relative effect of ${S_\text{1}}$ on $E_\text{x,n}$ is maximum at highest $K$ and lowest ${S_\text{rh}}$ at P$_2$ (\fig\ref{fig:9}). For instance, $E_\text{x,n}$ changes for (P$_1$, P$_2$, P$_3$, P$_4$) by $-2.57\%$ (from $0.5274$ to $0.5138$), (from $3.4822\times10^{-11}$ to $5.0003\times10^{-10}$), $4.88\%$ (from $0.4732$ to $0.4963$), 0\%, respectively with enhancing ${S_\text{1}}$ from 4 to 16 at $K=20$ and $ {S_\text{rh}}=0$ (refer \fig\ref{fig:9}). 

Further, $E_\text{x,n}$ enhances with increasing ${S_\text{rh}}$ but at higher ${S_\text{rh}}$ and lower $K$, it decreases with increasing ${S_\text{rh}}$, irrespective of ${S_\text{1}}$ (\fig\ref{fig:9}). It is because EDLs cover most of the cross-section area of the device at higher ${S_\text{rh}}$ and lower $K$, which impedes the flow of excess ions in the fluid. Maximum variation in $E_\text{x,n}$ with ${S_\text{rh}}$ is recorded at lowest ${S_\text{1}}$ and highest $K$ at P$_2$ (\fig\ref{fig:9}). For instance, increment in the values of $E_\text{x,n}$ is noted as 180.03\% (from 0.5274 to 1.4769), (from $3.48221\times10^{-11}$ to 1.9856), 219.23\% (from 0.4732 to 1.5107), 0\% for (P$_1$, P$_2$, P$_3$, P$_4$), respectively with increasing charge-heterogeneity ($0\le {S_\text{rh}}\le 2$) at weak EVF ($K=20$ and ${S_\text{1}}=4$) condition (refer \fig\ref{fig:9}a). 
\begin{figure}[b!]
	\centering\includegraphics[width=1\linewidth]{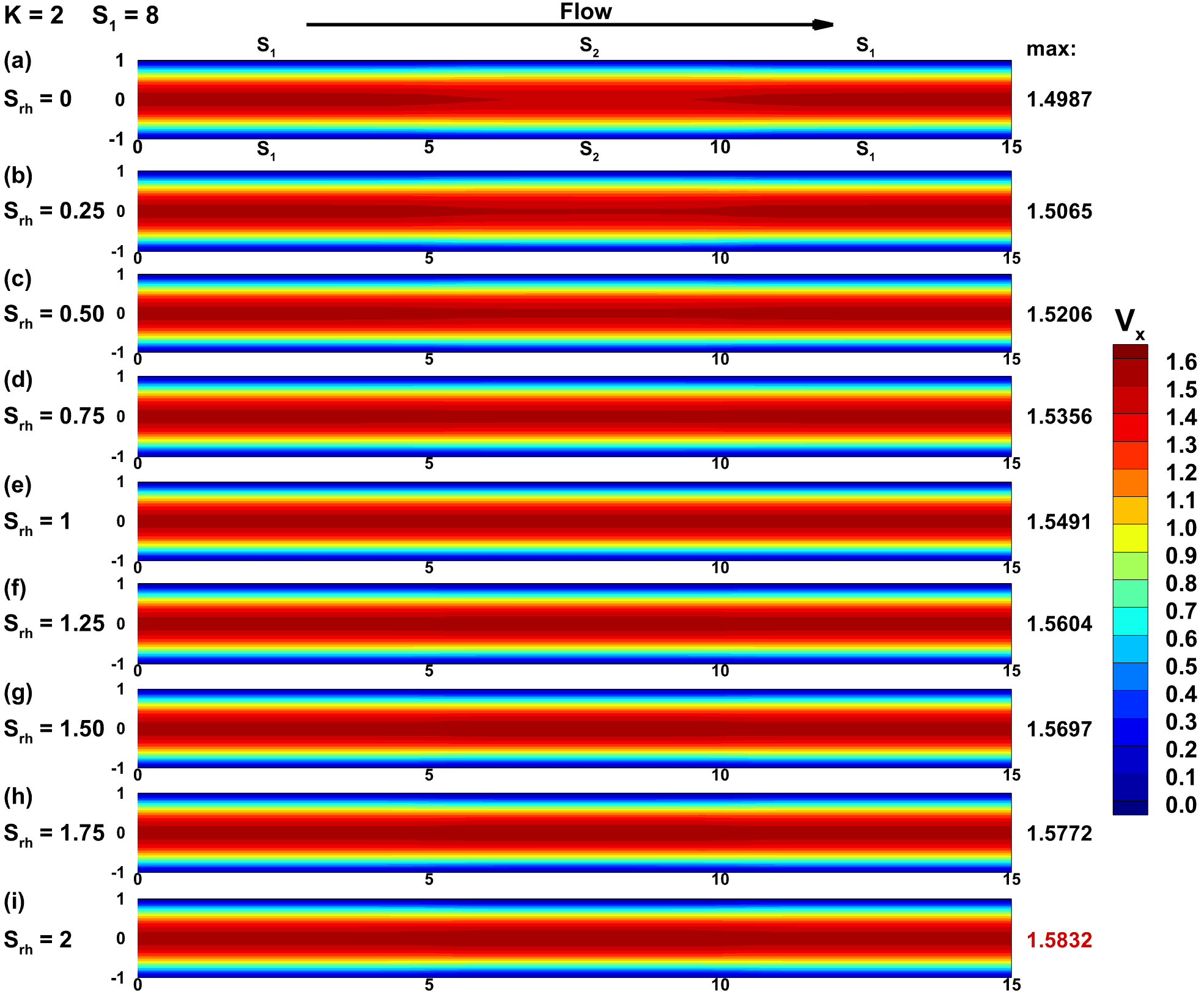}
	\caption{Influence of charge-heterogeneity ($0\le \mathit{S_\text{rh}}\le 2$) on velocity (${V}_{x}$) field distribution in the uniform microfluidic device at $K=2$ and $\mathit{S_\text{1}}=8$.}
	\label{fig:8v1}
\end{figure} 
Thus, it is noted that maximum variation in $E_\text{x,n}$ with governing parameters ($K$, ${S_\text{1}}$, ${S_\text{rh}}$) are obtained at P$_2$ as compared to the other locations (P$_1$, P$_3$, P$_4$) of device. It attributes that point P$_2$ is maximum affected by relative variation of surface charge densities of heterogeneous (${S_\text{2}}$) and upstream/downstream ($ {S_\text{1}}$) section. It is because $n^\ast$ depicted significant variation with dimensionless parameters ($K$, ${S_\text{1}}$, ${S_\text{rh}}$) at point P$_2$ as discussed in section \ref{sec:charge} (refer \fig\ref{fig:7}), thus, from \eqn(\ref{eq:2}) $E_\text{x}\propto n^\ast$ and hence $E_\text{x,n}$ have shown maximum variation at point P$_2$ (\fig\ref{fig:9}).

The above ensuing sections have shown the complex dependence of total potential, excess charge, and induced electric field strength on the dimensionless flow governing parameters ($K$, ${S_\text{1}}$, ${S_\text{rh}}$), corresponding influences of governing parameters on the {velocity and} pressure fields have been analyzed in the next sections. 
\subsection{Velocity ($\mathbf{V}$) field}
\begin{figure}[b!]
	\centering\includegraphics[width=1\linewidth]{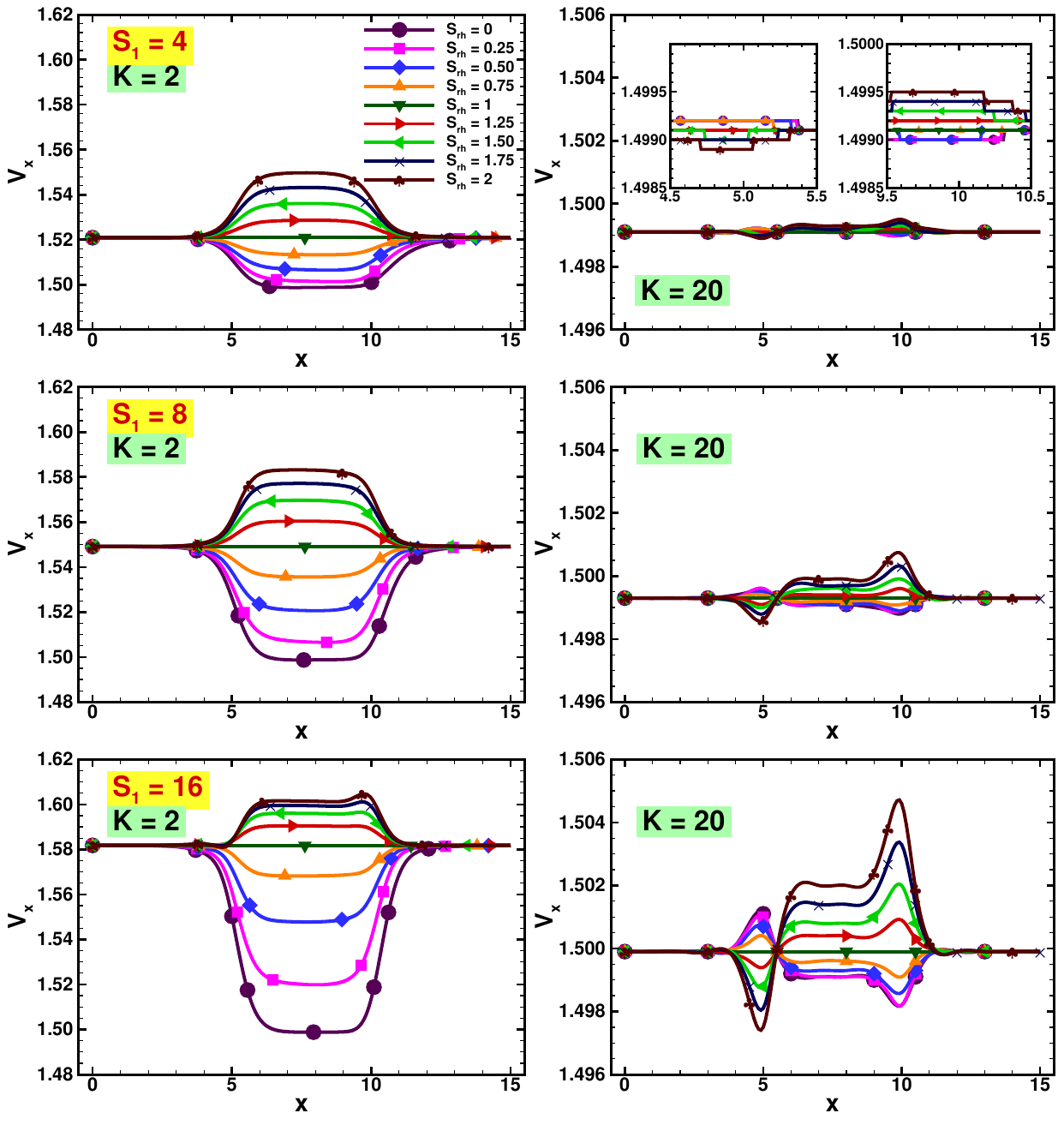}
	\caption{Velocity (${V}_{x}$) profiles on the centreline (P$_0$ to P$_4$; \fig\ref{fig:1}) of the device as a function of the dimensionless parameters ($K$, ${S_\text{1}}$, ${S_\text{rh}}$).}
	\label{fig:5vp4}
\end{figure}

{\fig\ref{fig:8v1} depicts the distribution of the axial component of velocity (${V}_{x}$) field in the considered microfluidic device for $0\le {S_\text{rh}}\le 2$, $K=2$ and ${S_\text{1}}=8$; qualitatively similar profiles are observed for the other conditions (\tab\ref{tab:1b}), and thus not presented here. As expected, qualitative nature of the velocity contour profiles are similar to that in pressure-drive flow through electrically neutral (nEVF, $K=\infty$ or ${S_\text{k}}=0$), homogeneously charged (${S_\text{rh}}=1$) microfluidic device. }
\begin{figure}[b!]
	\centering\includegraphics[width=0.97\linewidth]{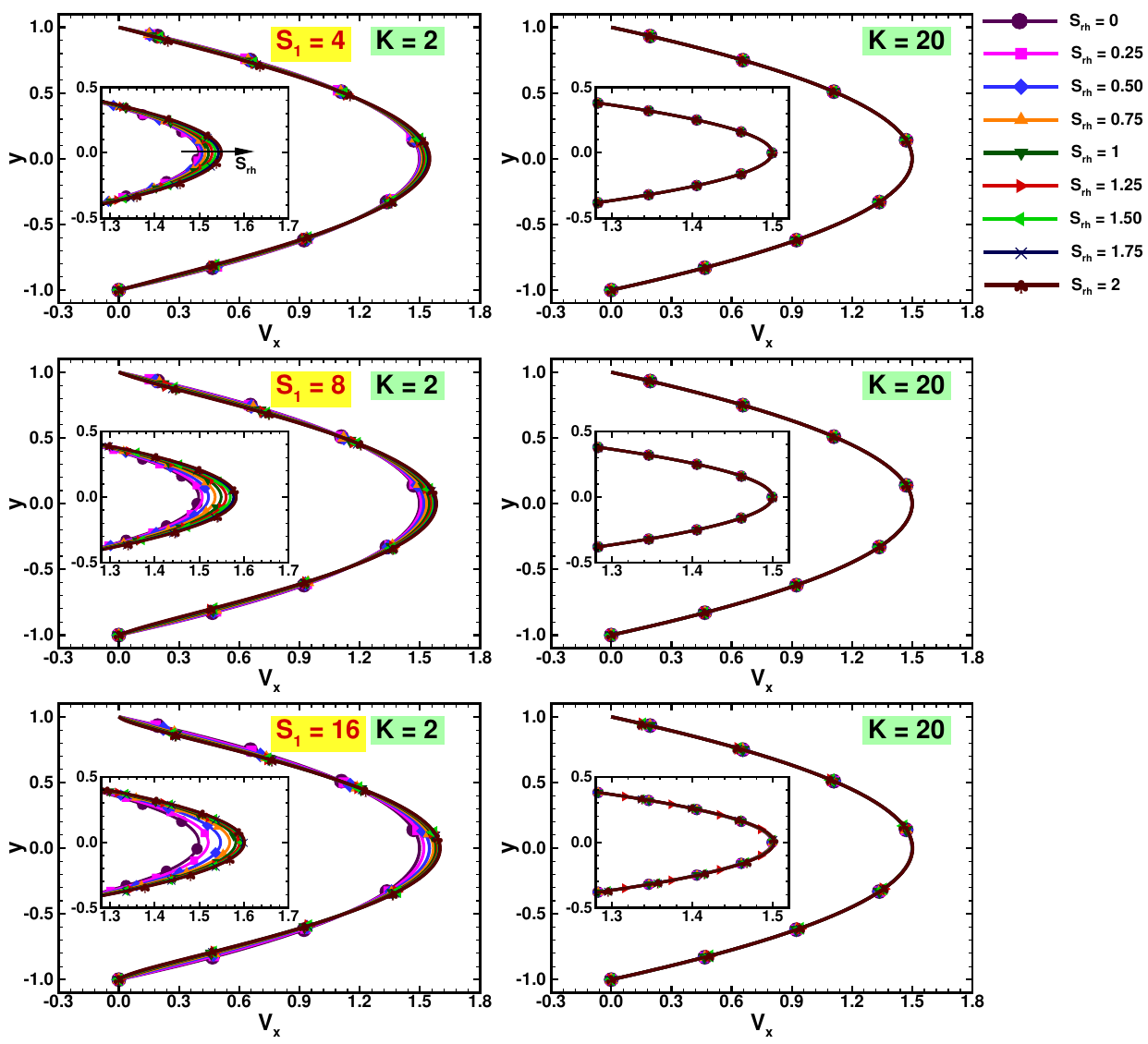}
	\caption{Velocity (${V}_{x}$) variation on the mid-plane ($L/2, y$) of microfluidic device as a function of the dimensionless parameters ($K$, ${S_\text{1}}$, ${S_\text{rh}}$).}
	\label{fig:5vc4}
\end{figure} 
{Subsequently, quantitative influence of the flow governing parameters ($K$, ${S_\text{1}}$, ${S_\text{rh}}$) on the velocity (${V}_{x}$) profiles are depicted on the mid-plane ($L/2, y$) and on the centerline (P$_0$ to P$_4$; \fig\ref{fig:1}) of the heterogeneously charged microfluidic device in \figs\ref{fig:5vc4} and \ref{fig:5vp4}, respectively. The velocity (${V}_{x}$) varies from zero (at walls, $y=\pm W$) to maximum (at the centerline, $y=0$) in the microchannel due to no-slip condition on the channel walls (refer \figs\ref{fig:8v1} and \ref{fig:5vc4}). The maximum velocity ($V_\text{max}={V}_{x,\text{max}}$) has shown enhancement with increasing ${S_\text{rh}}$, irrespective of ${S_\text{1}}$ and $K$  (refer \fig\ref{fig:5vc4}). For instance, $V_\text{max}$ is noted to enhance from 1.4987 to 1.5832 with ${S_\text{rh}}$ variation from 0 to 2 at $K=2$ and $\mathit{S_\text{1}}=8$ (refer \fig\ref{fig:8v1}). 
Broadly, the increment in maximum velocity ($V_\text{max}$) is recorded with increasing $S_\text{1}$, $S_\text{rh}$ and decreasing $K$ (or thickening of EDL). It is because, with increasing both $S_\text{k}$ and $S_\text{rh}$ and decreasing $K$, increased electrostatic interaction, i.e., ionic rearrangement and development of EDL retards the flow near the charged walls (refer \fig\ref{fig:5vc4}), and thus, the maximum velocity ($V_\text{max}$) on the centreline of the device  enhances  for the fixed volumetric flow rate ($Q$).}

{Subsequently, velocity (${V}_{x}$) profiles on the centreline (P$_0$ to P$_4$; \fig\ref{fig:1}) of the device as a function of the dimensionless flow governing parameters ($K$, ${S_\text{1}}$, ${S_\text{rh}}$) are analyzed in \fig\ref{fig:5vp4}. 
As expected, the velocity is uniform in all the three  (upstream, heterogeneous, and downstream) sections of the microfluidic geometry for homogeneously charged (${S_\text{rh}}=1$, $K$, ${S_\text{1}}$) conditions. However, ${V}_{x}$ is observed to be smaller for ${S_\text{rh}}<1$ (and larger for ${S_\text{rh}}>1$) in the heterogeneous section ($5\le x\le 10$) as compared with the upstream/downstream sections ($5>x>10$) of the device under otherwise identical conditions ($K$, ${S_\text{1}}$), as shown in \fig\ref{fig:5vp4}. While there is a smooth variation in the velocity at the intersections (i.e., lines $a$ and $b$, refer \fig\ref{fig:1}) at the lower values of $K$ and ${S_\text{1}}$, the complex nature is observed at these intersections in \fig\ref{fig:5vp4} with increasing values of the parameters ($K$, ${S_\text{1}}$). Notably, the charge heterogeneity (${S_\text{rh}}<1$) influences on the velocity field are enhances significantly with enhanced charge (${S_\text{1}}$) on the walls. 
Furthermore, quantitative influence of flow parameters ($K$, ${S_\text{1}}$, ${S_\text{rh}}$) on  $V_\text{max}$ is summarized in  \tab\ref{tab:v4}. The $V_\text{max}$ increases with decreasing $K$ and enhancing both ${S_\text{1}}$ and ${S_\text{rh}}$  (refer \fig\ref{fig:5vp4} and \tab\ref{tab:v4}).  
For instance, $V_\text{max}$ changes for (${S_\text{1}}=4$, 8, 16) by ($1.39\%$, $3.27\%$, $5.45\%$) and ($-0.06\%$, $-0.05\%$, $-0.01\%$) respectively at $K=2$ and 20 under the homogeneously charged (${S_\text{rh}}=1$) condition with respect to nEVF ($K=\infty$, ${S_\text{k}=0}$) condition; the corresponding changes under the heterogeneously charged ($\mathit{S_\text{rh}}\neq 1$) condition are ($-0.09\%$, $-0.09\%$, $-0.08\%$) and ($-0.07\%$, $-0.08\%$, $-0.13\%$) at $K=2$ and 20 for $\mathit{S_\text{rh}}=0$, and ($3.32\%$, $5.55\%$, $6.98\%$) and ($-0.03\%$, $0.05\%$, $0.31\%$) at $K=2$ and 20 for $\mathit{S_\text{rh}}=2$, respectively (refer \fig\ref{fig:5vp4} and \tab\ref{tab:v4}).  
Furthermore, $V_\text{max}$ variation is observed for (${S_\text{1}}=4$, 8, 16) by ($-1.46\%$, $-3.25\%$, $-5.25\%$) and ($-0.01\%$, $-0.03\%$, $-0.12\%$) at $K=2$ and 20 as $\mathit{S_\text{rh}}$ reduced from 1 (homogeneous) to 0 (heterogeneous); the corresponding variations are observed as ($1.90\%$, $2.20\%$, $1.45\%$) and ($0.03\%$, $0.10\%$, $0.32\%$) at $K=2$ and 20 as $\mathit{S_\text{rh}}$ increased from 1 (homogeneous) to 2 (heterogeneous).  Overall, $V_\text{max}$ changes for (${S_\text{1}}=4$, 8, 16) by ($3.41\%$, $5.64\%$, $7.07\%$) and ($0.03\%$, $0.13\%$, $0.44\%$) at $K=2$ and 20 as $\mathit{S_\text{rh}}$ increased from 0 to 2 (refer \tab\ref{tab:v4}).}
\begin{table}[t!]
	\centering
	\caption{Maximum velocity ($V_\text{max}$) on the centreline of charged microfluidic device as a function of governing parameters ($K$, $S_\text{1}$, and $S_\text{rh}$). Largest values of $V_\text{max}$ are underlined for each $K$ and $S_\text{1}$. }\label{tab:v4}
	\scalebox{0.85}
	{
		\begin{tabular}{|r|r|c|c|c|c|c|c|c|c|c|}
			\hline
			$S_\text{1}$	&	$K$	&	\multicolumn{9}{c|}{$V_\text{max}$}	\\\cline{3-11}
			&		&	$S_\text{rh}=0$	&	$S_\text{rh}=0.25$	& $S_\text{rh}=0.50$ &	$S_\text{rh}=0.75$ & $S_\text{rh}=1$ & $S_\text{rh}=1.25$ & $S_\text{rh}=1.50$ & $S_\text{rh}=1.75$ & $S_\text{rh}=2$ \\\hline
			\multicolumn{2}{|c|}{nEVF}    & 	\multicolumn{9}{c|}{1.5}	 \\\hline
			4	&	2	&	1.4987	&	1.5014	&	1.5065	&	1.5133	&	1.5209	&	1.5286	&	1.5361	&	1.5432	&	\underline{1.5498}	\\
			&	20	&	1.4990	&	1.4990	&	1.4990	&	1.4991	&	1.4991	&	1.4992	&	1.4993	&	1.4994	&	\underline{1.4995}	\\\hline
			8	&	2	&	1.4987	&	1.5065	&	1.5206	&	1.5356	&	1.5491	&	1.5604	&	1.5697	&	1.5772	&	\underline{1.5832}	\\
			&	20	&	1.4988	&	1.4988	&	1.4989	&	1.4991	&	1.4993	&	1.4996	&	1.4999	&	1.5003	&	\underline{1.5008}	\\\hline
			16	&	2	&	1.4988	&	1.5199	&	1.5477	&	1.5682	&	1.5818	&	1.5905	&	1.5965	&	1.6011	&	\underline{1.6047}	\\
			&	20	&	1.4981	&	1.4982	&	1.4985	&	1.4991	&	1.4999	&	1.5009	&	1.5021	&	1.5034	&	\underline{1.5047}	\\\hline
		\end{tabular}
	}
\end{table}
\subsection{Pressure ($P$) distribution}
\label{sec:pressure}
\begin{figure}[!b]
	\centering\includegraphics[width=1\linewidth]{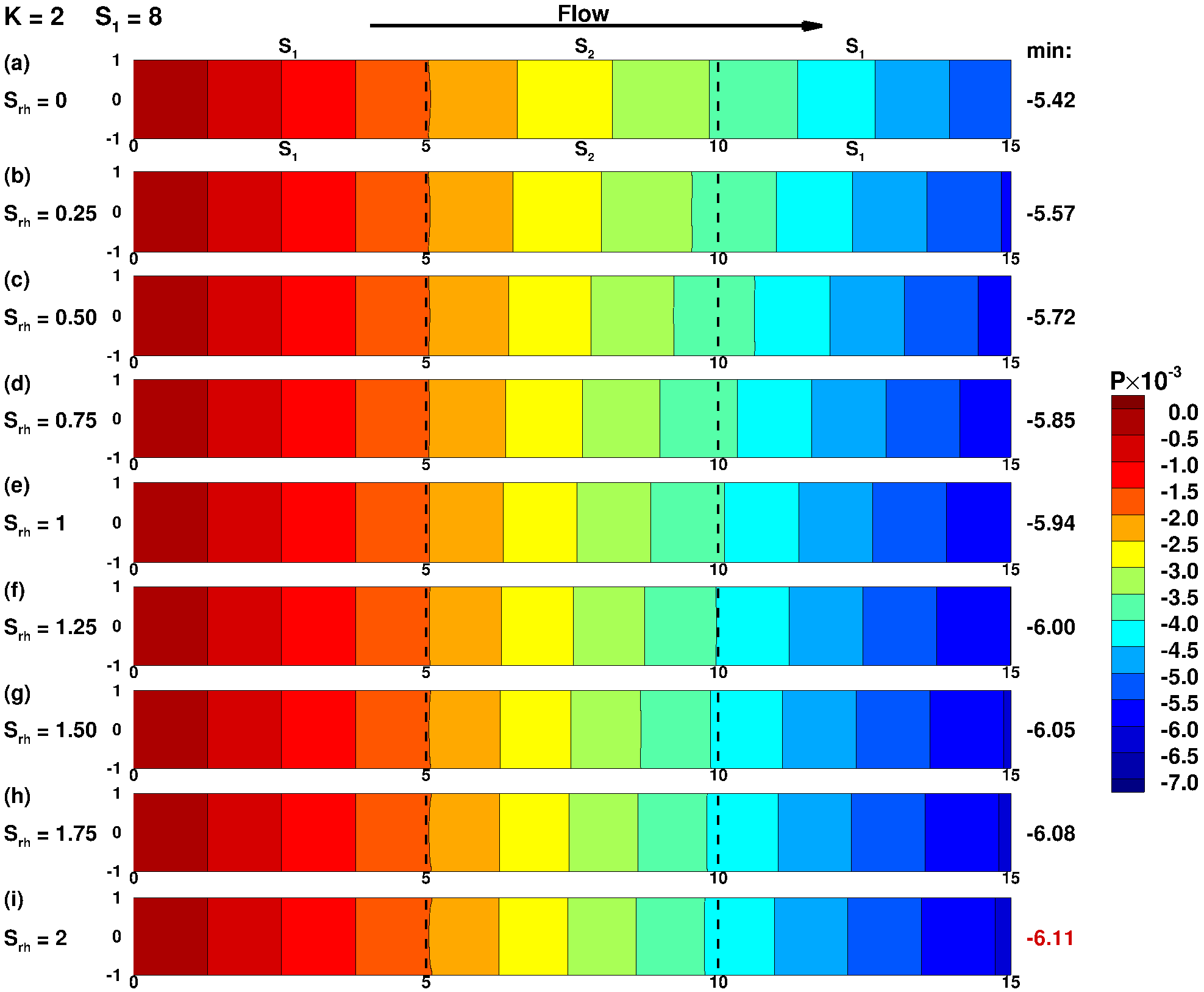}
	\caption{Pressure ($P^\ast=P\times10^{-3}$) distribution in a heterogeneously charged ($0\le {S_\text{rh}}\le 2$) device for a fixed condition (${S_\text{1}}=8$ and $K=2$).}
	\label{fig:10}
\end{figure}

%
\fig\ref{fig:10} depicts the pressure ($P^\ast= P\times10^{-3}$) distribution in a heterogeneously charged ($0\le {S_\text{rh}}\le 2$) device at fixed ${S_\text{1}}=8$ and $K=2$; other conditions ($K$, $ {S_\text{1}}$, $ {S_\text{rh}}$; \tab\ref{tab:1b}) show qualitatively similar contour profiles which are not shown here. As expected, the pressure reduces along the length ($0\le x\le L$) of the microfluidic device due to increased resistance by both hydrodynamic and electrostatic forces (\fig\ref{fig:10}). 
Further, it decreases with increasing ${S_\text{rh}}$ (\fig\ref{fig:10}) due to strengthening in the charge attractive force close to the device walls, which imposes an extra resistance on the PDF flow of liquid in the device. For instance, minimum value of pressure drop ($\Delta P^\ast$) is noted as $-6.11$ at ${S_\text{rh}}=2$ for $K=2$ and ${S_\text{1}}=8$ (\fig\ref{fig:10}i). However, overall minimum value of $\Delta P^\ast$ is recorded as $-6.36$ at ${S_\text{rh}}=1.5$, $K=2$, and ${S_\text{1}}=16$. In general, the pressure gradient increases significantly in the heterogeneous section, followed by relatively less enhancement in the downstream section when ${S_\text{rh}}$ changes from 0 to 2 (\fig\ref{fig:10}). It is due to the charge-heterogeneity on the device walls that imposes non-uniform additional resistance on the liquid flow by charge-attractive force along the device.

\begin{figure}[!b]
	\centering\includegraphics[width=1\linewidth]{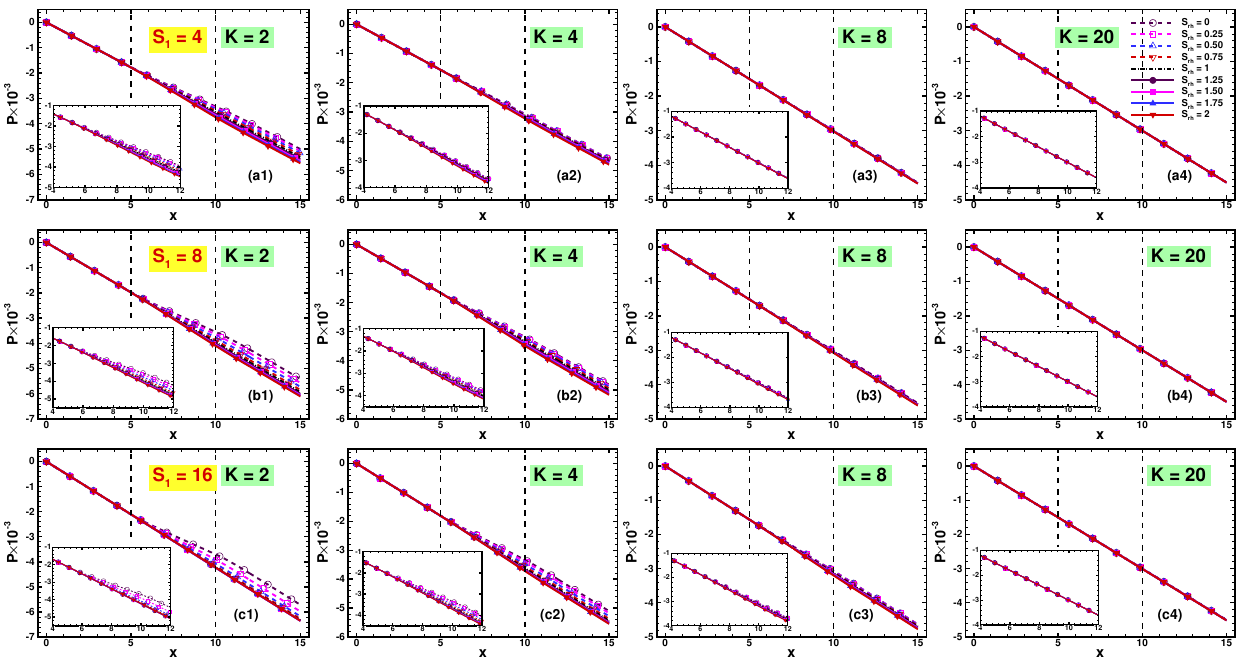}
	\caption{Pressure ($P^\ast$) variation on the centreline (P$_0$ to P$_4$; \fig\ref{fig:1}) of heterogeneously charged microfluidic device for dimensionless parameters ($K$, ${S_\text{1}}$, ${S_\text{rh}}$; \tab\ref{tab:1b}).}
	\label{fig:11}
\end{figure}

Further, \fig\ref{fig:11} shows pressure ($P^\ast$) variation on the centreline (P$_0$ to P$_4$; \fig\ref{fig:1}) of the device for governing parameters ($K$, ${S_\text{1}}$, ${S_\text{rh}}$; \tab\ref{tab:1b}). The pressure ($P$) decreases along the length of the device, irrespective of the flow conditions ($K$, ${S_\text{1}}$, ${S_\text{rh}}$) (\fig\ref{fig:11}). In upstream ($0\le x\le 5$) and downstream ($10\le x\le 15$) sections, the pressure gradient ($\Delta P$) varies uniformly along the length of the device, irrespective of the flow conditions. In heterogeneous ($5\le x\le 10$) section, the pressure gradient ($\Delta P$) increases with  increasing ${S_\text{rh}}$ form 0 to 2 at fixed $K$ and ${S_\text{1}}$ (\fig\ref{fig:11}). The pressure ($P$) decreases with decreasing $K$ and with increasing both ${S_\text{1}}$ and ${S_\text{rh}}$ (\fig\ref{fig:11}). Maximum variation in pressure drop ($|\Delta P|$) is obtained as 29.07\% when $K$ varies from 2 to 20 at ${S_\text{1}}=16$ and ${S_\text{rh}}=2$ (\fig\ref{fig:11}c).
%

Subsequently, \tab\ref{tab:1} comprises the pressure drop ($|\Delta P|$) on the centreline (P$_0$ to P$_4$; \fig\ref{fig:1}) of the heterogeneously charged device as a function of the flow parameters ($K$, ${S_\text{1}}$, $ {S_\text{rh}}$). Maximum values of $|\Delta P|$ for $0\le {S_\text{rh}}\le 2$ at each ${S_\text{1}}$ and $K$ are also highlighted with bold data. 
The magnitude of pressure drop ($|\Delta P|$) decreases with increasing $K$, irrespective of ${S_\text{1}}$ and ${S_\text{rh}}$ (\tab\ref{tab:1}). The variation in $|\Delta P|$ with $K$ is maximum at highest ${S_\text{1}}=16$ and ${S_\text{rh}}=2$ (\tab\ref{tab:1}). For instance, $|\Delta P|$ reduces by (10.02\%, 15.57\%, 19.09\%) at ${S_\text{1}}=4$ and (21.21\%, 28.78\%, 28.94\%)  at ${S_\text{1}}=16$ for (${S_\text{rh}}=0$, $1$, $2$), respectively when $K$ varies from 2 to 20 (refer \tab\ref{tab:1}). A maximum change in $|\Delta P|$ with ${S_\text{1}}$ is obtained at ${S_\text{rh}}=1$ and lowest $K=2$ (\tab\ref{tab:1}). For instance, $|\Delta P|$ increases at $K=2$ by (14.35\%, 18.77\%, 14.22\%) and at $K=20$ by (0.12\%, 0.18\%, 0.32\%) for (${S_\text{rh}}=0$, $1$, $2$), respectively with increasing ${S_\text{1}}$ from $4$ to $16$ (refer \tab\ref{tab:1}). 
%
Further, the impact of ${S_\text{rh}}$ on $|\Delta P|$ is observed maximum at ${S_\text{1}}=8$ and lowest $K=2$ (\tab\ref{tab:1}). For instance, $|\Delta P|$ reduces with decreasing ${S_\text{rh}}$ from 1 to 0 by (6.17\%, 8.81\%, 9.66\%) at $K=2$ and (0\%, 0.01\%, 0.06\%) at $K=20$ for (${S_\text{1}}=4$, 8, 16), respectively; On the other hand, increment in $|\Delta P|$ is noted with increasing ${S_\text{rh}}$ from 1 to 2 as (4.37\%, 2.83\%, 0.38\%) and (0.01\%, 0.05\%, 0.16\%) at $K=2$ and 20 for (${S_\text{1}}=4$, 8, 16), respectively. Overall enhancement in $|\Delta P|$ for (${S_\text{1}}=4$, 8, 16) is recorded as (11.24\%, 12.77\%, 11.11\%) at $K=2$ and (0.02\%, 0.06\%, 0.21\%) at $K=20$ with increasing ${S_\text{rh}}$ from 0 to 2 (refer \tab\ref{tab:1}). In general, $|\Delta P|$ increases with increasing ${S_\text{1}}$ and ${S_\text{rh}}$. It is because strengthening in the charge attractive force with increasing both ${S_\text{rh}}$ and ${S_\text{1}}$, which increases additional resistance in the pressure-driven flow; thus, $|\Delta P|$ increases from \eqn(\ref{eq:9}) with increasing additional resistance imposed by electrical force ($\myvec{F_\text{e}}$) (refer \fig\ref{fig:11} and \tab\ref{tab:1}).
\begin{figure}[!b]
	\centering\includegraphics[width=1\linewidth]{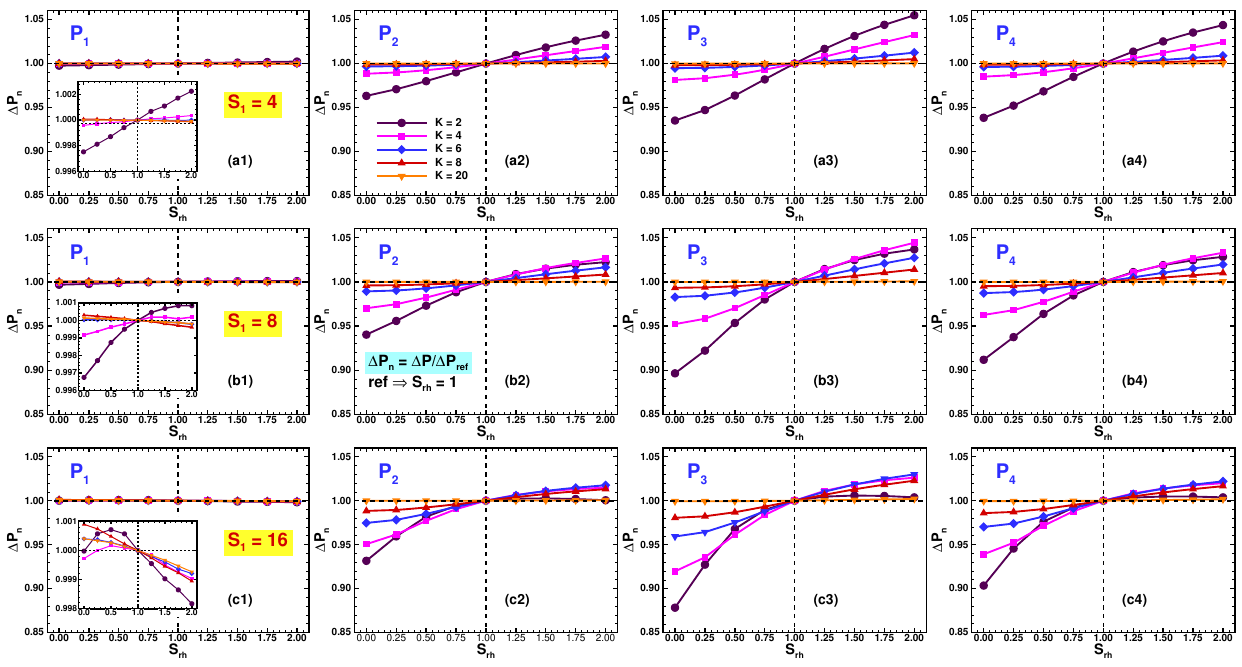}
	\caption{Normalized pressure drop ($\Delta P_\text{n}$, \eqn\ref{eq:11a}) variation with ${S_\text{rh}}$ on centreline points (P$_\text{j}$, \fig\ref{fig:1}) of heterogeneously charged microfluidic device for $2\le K\le 20$ and $4\le {S_\text{1}}\le 16$.}
	\label{fig:12}
\end{figure} 
%

Furthermore, \fig\ref{fig:12} shows normalized pressure drop ($\Delta P_\text{n}$, \eqn\ref{eq:11a}) variation with ${S_\text{rh}}$ on the centreline points (P$_\text{j}$, \fig\ref{fig:1}) of the device for $2\le K\le 20$ and $4\le {S_\text{1}}\le 16$. The normalized values depict complex dependency on governing parameters ($K$, ${S_\text{1}}$, and ${S_\text{rh}}$) at centreline points of device. The $\Delta P_\text{n}$ enhances with increasing $K$ for ${S_\text{rh}}<1$. However, it decreases with increasing $K$ for ${S_\text{rh}}>1$, followed by reverse trends at higher ${S_\text{1}}$ and lower $K$ (\fig\ref{fig:12}). Maximum variation in $\Delta P_\text{n}$ with $K$ is obtained at lowest ${S_\text{rh}}$ and highest ${S_\text{1}}$ at P$_3$ (\fig\ref{fig:12}). For instance, $\Delta P_\text{n}$ enhances by (0.04\%, 7.33\%, 13.78\%, 10.63\%) for (P$_1$, P$_2$, P$_3$, P$_4$), respectively with increasing $K$ from 2 to 20 at ${S_\text{1}}=16$ and ${S_\text{rh}}=0$ (refer \fig\ref{fig:12}c). The $\Delta P_\text{n}$ decreases with increasing ${S_\text{1}}$ for all centreline points of device; maximum variation in $\Delta P_\text{n}$ with ${S_\text{1}}$ is observed at lowest $K$ and highest ${S_\text{rh}}$ at P$_3$ (\fig\ref{fig:12}). For instance, $\Delta P_\text{n}$ reduces when ${S_\text{1}}$ varies from 4 to 16 by (0.41\%, 3.13\%, 4.84\%, 3.83\%) for (P$_1$, P$_2$, P$_3$, P$_4$), respectively at $K=2$ and ${S_\text{rh}}=2$ (refer \fig\ref{fig:12}). 
%
Further, $\Delta P_\text{n}$ enhances with increasing ${S_\text{rh}}$ but shows reverse trends at higher ${S_\text{rh}}$ and lower $K$ (\fig\ref{fig:12}). It is because EDL overlaps at higher ${S_\text{rh}}$ and lower $K$, which impedes excess ions flow downstream. The relative effect of ${S_\text{rh}}$ on $\Delta P_\text{n}$ is maximum at lowest $K=2$ and ${S_\text{1}}=8$ at P$_3$ (\fig\ref{fig:12}). For instance, increment in the values of $\Delta P_\text{n}$ are noted as (0.41\%, 8.71\%, 15.62\%, 12.76\%) for (P$_1$, P$_2$, P$_3$, P$_4$), respectively, with enhancing ${S_\text{rh}}$ from 0 to 2 at $K=2$ and ${S_\text{1}}=8$ (\fig\ref{fig:12}b). Thus, it is observed that $\Delta P_\text{n}$ varies maximally at P$_3$ than other centreline locations (P$_1$, P$_2$, P$_4$). It is because maximum variation of $\Delta U_\text{n}$ at P$_3$ as discussed in section \ref{sec:potential} (refer \fig\ref{fig:4}) imposes maximum variation in $P$ as $\nabla P\propto\nabla U$ (\eqn\ref{eq:9}) and, hence, $\Delta P_\text{n}$ at P$_3$ than other centreline points (P$_1$, P$_2$, P$_4$) of the microfluidic device (\fig\ref{fig:12}).
%
\subsection{Electroviscous correction factor ($Y$)}
\label{sec:ECF}
%
In the {pressure-driven} electrokinetic flows, streaming potential ($\phi$, \eqn\ref{eq:4a}) arises from the electric field strength ($E_\text{x}$) induced by the transport of excess charge through the microfluidic device. Streaming potential imposes an additional hydrodynamic resistance in the fluid flow depicted by the electrical force ($\myvec{F}_\text{e}$) in \eqn(\ref{eq:9}), which manifests the pressure drop ($\Delta P$) along the microchannel that is higher as compared to the pressure drop ($\Delta P_\text{0}$) for neutrally charged (nEVF, ${S_\text{k}}=0$ or $K=\infty$) walls, for a fixed flow rate ($Q$ m$^3$/s). The enhanced pressure drop is generally quantified in terms of the effective (or apparent) viscosity ($\mu_\text{eff}$), which is the viscosity needed to obtain the pressure drop ($\Delta P$) in the absence of electrical field (nEVF). This effect is commonly known as the `electroviscous effect' (EVE) \citep{davidson2007electroviscous,davidson2008electroviscous,bharti2008steady,dhakar2022electroviscous,dhakar2023cfd,dhakar2024influence,dhakar2024fmfp}. In the steady, laminar, low Reynolds number ($Re=10^{-2}$) flow, the non-linear advection term in the momentum equation (\eqn\ref{eq:9}) becomes negligible. Thus, the relative increment in the pressure drop ($\Delta P/\Delta P_0$) is attributed to the corresponding relative increased viscosity ($\mu_{\text{eff}}/\mu$), under otherwise identical conditions. Thus, \textit{electroviscous correction factor} ($Y$) is expressed as follows.
\begin{gather}
	Y=\frac{\mu_{\text{eff}}}{\mu}=\frac{\Delta P}{\Delta P_{\text{0}}}
	\label{eq:27}
\end{gather}
where $\mu$ is the physical viscosity of liquid.
\begin{figure}[!b]
	\centering
	\includegraphics[width=1\linewidth]{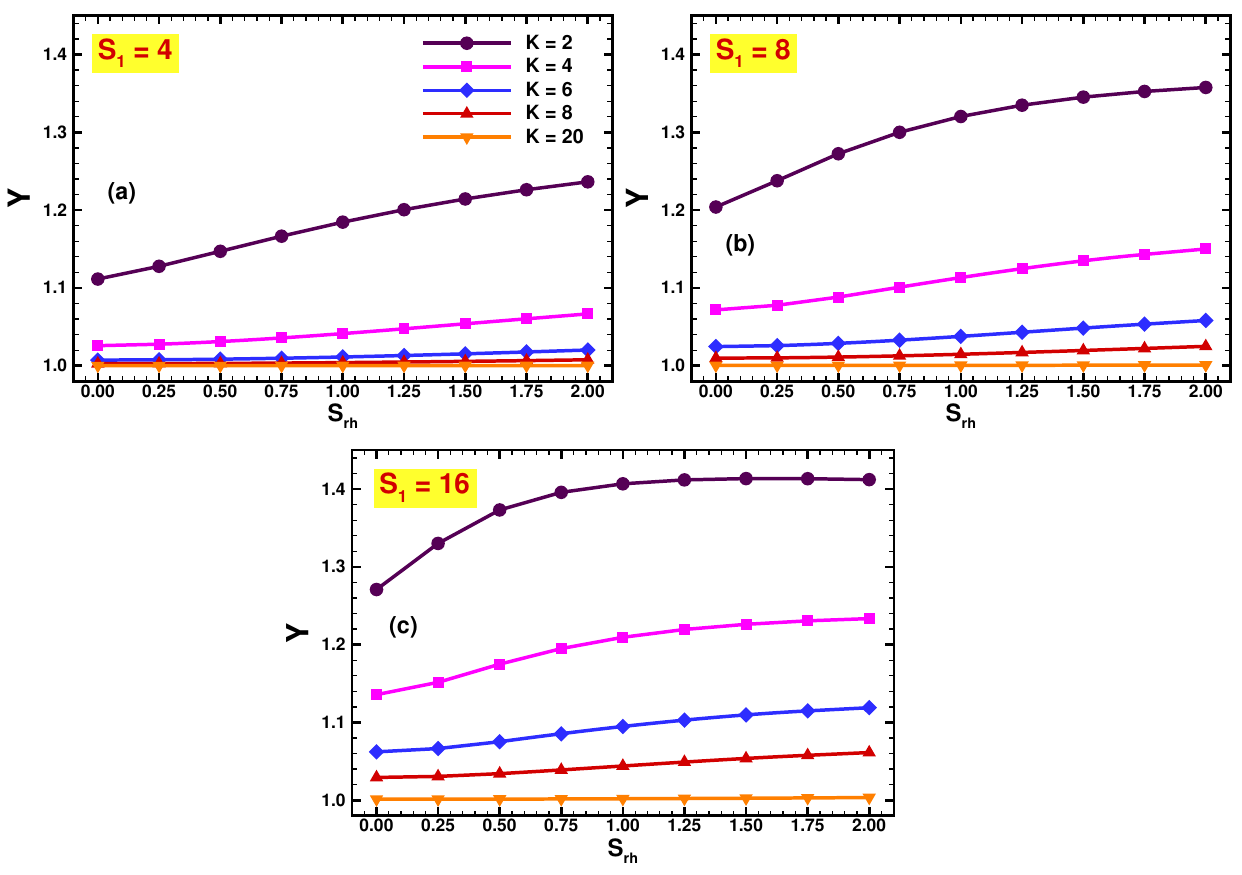}
	\caption{Electroviscous correction factor ($Y$) as a function of the flow parameters ($K$, ${S_\text{1}}$, ${S_\text{rh}}$).}
	\label{fig:13}
\end{figure} 
%

\fig\ref{fig:13} depicts the electroviscous correction factor ($Y$) variation with dimensionless parameters ($K$, ${S_\text{1}}$, ${S_\text{rh}}$; \tab\ref{tab:1b}). Electroviscous effects are absent when $Y=1$ and become stronger when $Y$ exceeds unity. It has shown complex dependency on the flow parameters ($K$, ${S_\text{1}}$, ${S_\text{rh}}$). In general, $Y$ increases with decreasing $K$, and with increasing ${S_\text{1}}$ and ${S_\text{rh}}$ (\fig\ref{fig:13}). It is due to an increment in the electrostatic force close to walls, which increases pressure drop as discussed in the section \ref{sec:pressure} and enhances the correction factor (\eqn\ref{eq:27}). Further, increment in $Y$ with ${S_\text{rh}}$ is significant for ${S_\text{rh}}<1$ (at smaller ${S_\text{1}}$) followed by relatively small for ${S_\text{rh}}>1$ (at higher ${S_\text{1}}$), irrespective of $K$. It is because at higher ${S_\text{rh}}$ and ${S_\text{1}}$, stronger electrostatic force retards the excess ions flow in the device (\fig\ref{fig:13}). For instance, $Y$ maximally increases by 40.98\% (at ${S_\text{1}}=16$, ${S_\text{rh}}=1.5$, $20\ge K\ge 2$), 19.72\% (at $K=2$, ${S_\text{rh}}=0.5$, $4\le S_\text{1}\le 16$), 12.77\% (at $K=2$, ${S_\text{1}}=8$, $0\le S_\text{rh}\le 2$). Overall increment in $Y$ is noted as 41.35\% at $K=2$, ${S_\text{1}}=16$, and ${S_\text{rh}}=1.5$ relative to nEVF (refer \fig\ref{fig:13}). 
Thus, charge-heterogeneity enhances the electroviscous effects in the microfluidic device. It enables the use of present numerical results for designing efficient and reliable microfluidic devices to control mixing efficiency and heat and mass transfer rates of the processes.
%

The predictive correlation depicting the functional dependence of the electroviscous correction factor ($Y$, \fig\ref{fig:13}) on the flow governing parameters ($K$, ${S_\text{1}}$, ${S_\text{rh}}$) is expressed as follows. 
\begin{numcases}
	{Y=}
	B_1 + (B_2+B_4X)X+ (B_3 + B_5  {S_\text{rh}}) {S_\text{rh}}+ B_6 X {S_\text{rh}}, & or \label{eq:y18}\\
	\exp(C_1 + C_2 X+ C_3  {S_\text{rh}}+C_4  {S_\text{1}} + C_5 X {S_\text{1}}+C_6X {S_\text{rh}}) & \label{eq:y19}
\end{numcases}
\begin{gather}
	\text{where}\qquad 
	B_{\text{i}} = \sum_{{j}=1}^3 N_{\text{ij}}  {S_\text{1}}^{({j}-1)},  \quad X=K^{-1}, \quad 1\le i\le 6 \nonumber
\end{gather}
The correlation coefficients ($B_{\text{i}}$, $C_{\text{i}}$) are statistically obtained, using 135 data points over the given ranges of conditions (\tab\ref{tab:1b}), by performing the non-linear regression analysis using DataFit (trial version) with ($\delta_{\text{min}}$, $\delta_{\text{max}}$, $\delta_{\text{avg}}$, $R^2$) as ($-3.26\%, 2.91\%, 0.10\%, 98.86\%$) for \eqn(\ref{eq:y18}), and  ($-4.57\%, 4.40\%, -0.01\%, 96.36\%$)  for \eqn(\ref{eq:y19}) as follows.
\begin{gather*}
	N = \begin{bmatrix}
	   1.0463	&	-0.0091	&	9\times10^{-5} \\
   	   -0.4936	&	0.0767	&	0.0007 \\
	   -0.031	&	0.00439	&	9\times10^{-6} \\
	   0.6887	&	0.0509	&	-0.007 \\
	   0.0069	&	-0.0019	&	2\times10^{-5} \\
	   0.0961	&	0.0179	&	-0.001	
	\end{bmatrix} 
\qquad\text{and}\qquad 
	C = 
	\begin{bmatrix}
	-0.05	\\	0.248	\\	-0.0014	\\	0.0011	\\	0.0236	\\	0.1105
\end{bmatrix}^{T} 
\end{gather*}
%
\subsection{Pseudo-analytical model for pressure drop ($\Delta P$) prediction}
%

This section has developed a simple pseudo-analytical model to predict the pressure drop ($\Delta P$), obtained numerically and discussed in the section \ref{sec:pressure}, for their easy utilization in the systematic design of the relevant microfluidic applications. Earlier studies  \citep{davidson2007electroviscous,bharti2008steady,dhakar2022electroviscous,dhakar2023cfd,dhakar2024influence} have proposed the simple pseudo-analytical models to approximate the pressure drop in flow through symmetrically (${S_\text{r}}=1$) / asymmetrically (${S_\text{r}}\neq1$) and homogeneously (${S_\text{rh}}=1$) charged non-uniform ($d_\text{c}=0.25$) microfluidic devices. 
Based on a similar approach \citep{davidson2007electroviscous,bharti2008steady,dhakar2022electroviscous,dhakar2023cfd,dhakar2024influence}, a pseudo-analytical model has been developed to estimate the pressure drop in the symmetric ($1:1$) electrolyte liquid flow through heterogeneously positively charged (${S_\text{k}} \ge 0$, ${S_\text{rh}}\neq 1$) uniform ($d_\text{c}=1$) slit microfluidic device, by summing up the pressure drop in the individual  (i.e., i$^\text{th}$) sections of the device, as follows. 
\begin{gather}
\Delta P_{\text{m}}=\left(\sum_{i=u,h,d}\Delta P_{\text{i}}\right)
\label{eq:33}
\end{gather}
where subscripts `u', `h', and `d' denote the upstream, heterogeneous, and downstream sections, respectively. These sections individually represents the uniform slit of rectangular cross-section. 
Referring \eqn(\ref{eq:27}), which correlates the pressure drop ($\Delta P$) under EVF ($K$, ${S_\text{1}}$, ${S_\text{rh}}$) and nEVF (${S_\text{k}}=0$ or $K=\infty$) conditions,  \eqn(\ref{eq:33}) can be simplified as follows.
\begin{gather}
	\Delta P_{\text{m}}= \Gamma_\text{hr} \Delta P_{0,\text{m}} 
	\qquad \text{where}\qquad
	\Delta P_{\text{0,m}}=\left(\sum_{i=u,h,d}\Delta P_{\text{0,i}}\right)
	\label{eq:38a}
\end{gather}
where $\Gamma_\text{hr}$ is the correction coefficient to the pressure drop ($\Delta P_{0,\text{m}}$) under nEVF condition, accounting for the influence of the heterogeneously positively charged (${S_\text{k}} \ge 0$, ${S_\text{rh}}\neq 1$) device.
%
%

The pressure drop (${\Delta P}_{0,\text{i}}$)  in the laminar steady fully-developed flow of incompressible Newtonian liquid through uniform (i.e., i$^\text{th}$) sections of the slit device, under nEVF condition, is analytically estimated  \citep{davidson2007electroviscous,bharti2008steady,dhakar2022electroviscous,dhakar2023cfd,dhakar2024influence} by the \textit{Hagen-Poiseuille equation} as follows.
\begin{gather}
	\label{eq:35}
	{\Delta P}_{0,\text{i}}= \left(\frac{3}{Re}\right){L_{\text{i}}}
\end{gather}
%
%
%
Thus, a generalized simpler pseudo-analytical model to predict the pressure drop in symmetric ($1:1$) electrolyte liquid flow through heterogeneously positively charged (${S_\text{k}} \ge 0$, ${S_\text{rh}}\neq 1$) uniform ($d_\text{c}=1$) slit microchannel is expressed as follows.
\begin{gather}
\Delta P_{\text{m}} =
\left(\frac{3\Gamma_\text{hr}}{Re}\right)(L_{\text{u}} +  L_{\text{h}} + L_{\text{d}})
\label{eq:38}
\end{gather}
%
The correction coefficient ($\Gamma_\text{hr}$), appearing in \eqns(\ref{eq:38a}) and (\ref{eq:38}), is correlated with EVF parameters ($K$, ${S_\text{1}}$, ${S_\text{rh}}$) as follows.
\begin{numcases}
	{\Gamma_\text{hr}=}
	B_1 + (B_2+B_4 X)X + (B_3+B_5  {S_\text{rh}}) {S_\text{rh}} + B_6 X {S_\text{rh}}, & or \label{eq:39}\\
	\exp(C_1 + C_2 X + C_3  {S_\text{rh}} + C_4  {S_\text{1}} + C_5 X {S_\text{1}}+ C_6X {S_\text{rh}}) & \label{eq:39a}
\end{numcases}
\begin{gather}
	\text{where}\qquad 
	B_{\text{i}} = \sum_{{j}=1}^3 N_{\text{ij}}  {S_\text{1}}^{({j}-1)} \quad \text{and}\quad X=K^{-1}, \quad 1\le i\le 6 \nonumber
\end{gather}
The correlation coefficients ($B_{\text{i}}$, $C_{\text{i}}$) are statistically obtained, using 135 data points over the given ranges of conditions (\tab\ref{tab:1b}), by performing the non-linear regression analysis using DataFit (trial version) with ($\delta_{\text{min}}$, $\delta_{\text{max}}$, $\delta_{\text{avg}}$, $R^2$) as ($-3.26\%, 2.97\%, 0.10\%, 98.86\%$) for \eqn(\ref{eq:39}), and  ($-4.59\%, 4.59\%, -0.01\%, 96.38\%$)  for \eqn(\ref{eq:39a}) as follows.
\begin{gather}
	N = \begin{bmatrix}
     	1.0455	&	-0.0091	&	9\times10^{-5} \\
    	-0.4932	&	0.0767	&	0.0007  \\
    	-0.031	&	0.00439	&	9\times10^{-6} \\
    	0.6881	&	0.0494	&	-0.0069 \\
    	0.0069	&	-0.0019	&	2\times10^{-5} \\
    	0.096	&	0.0179	&	-0.001	
	\end{bmatrix} 
	\qquad\text{and}\qquad
		C = \begin{bmatrix}
		-0.0535	\\	0.249	\\	-0.0018	\\	0.0014	\\	0.0236	\\	0.1105
	\end{bmatrix} ^{T}
	\nonumber
\end{gather}
\begin{figure}[!tb]
	\centering
	\subfigure[]{\includegraphics[width=0.49\linewidth]{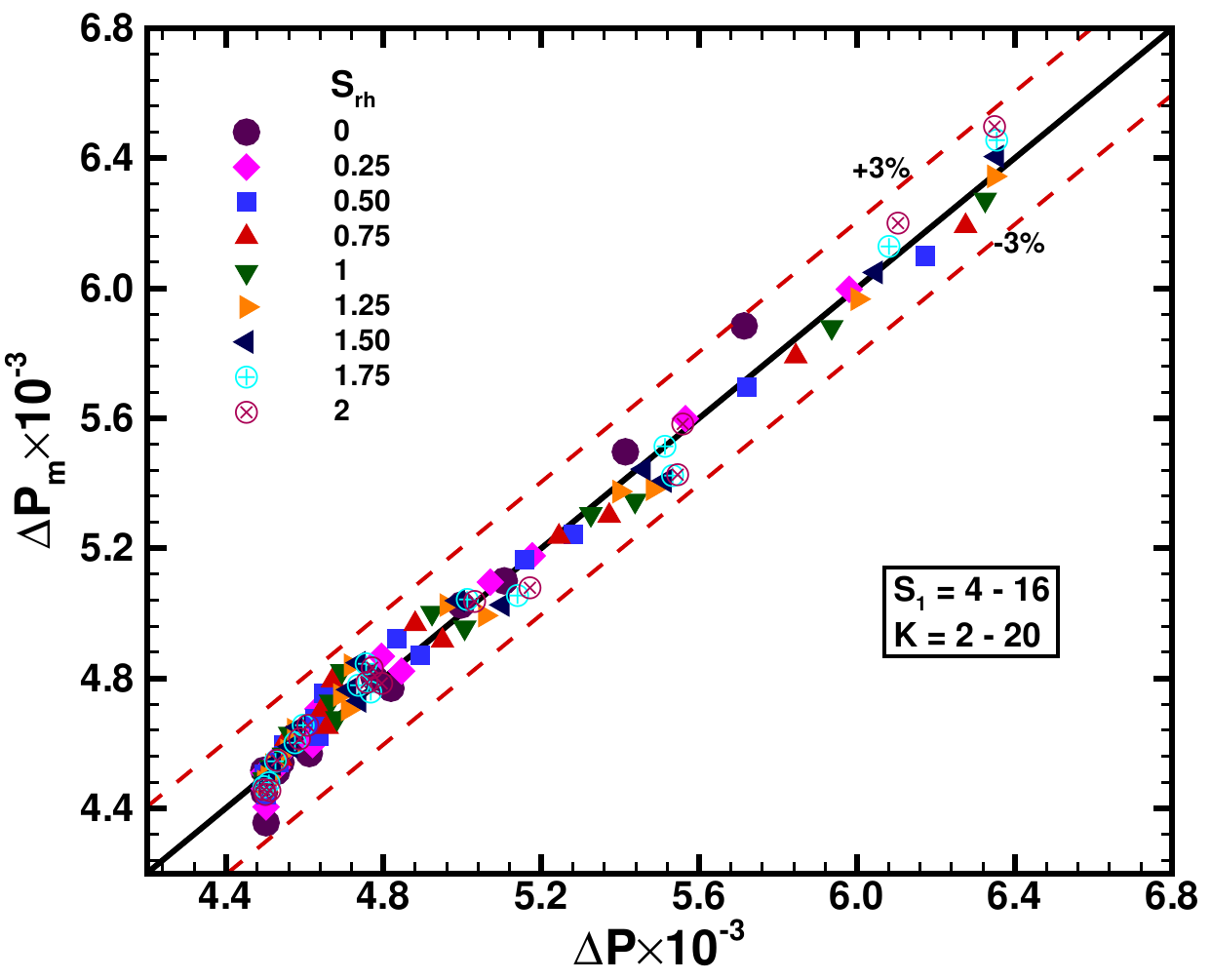}}
	\subfigure[]{\includegraphics[width=0.49\linewidth]{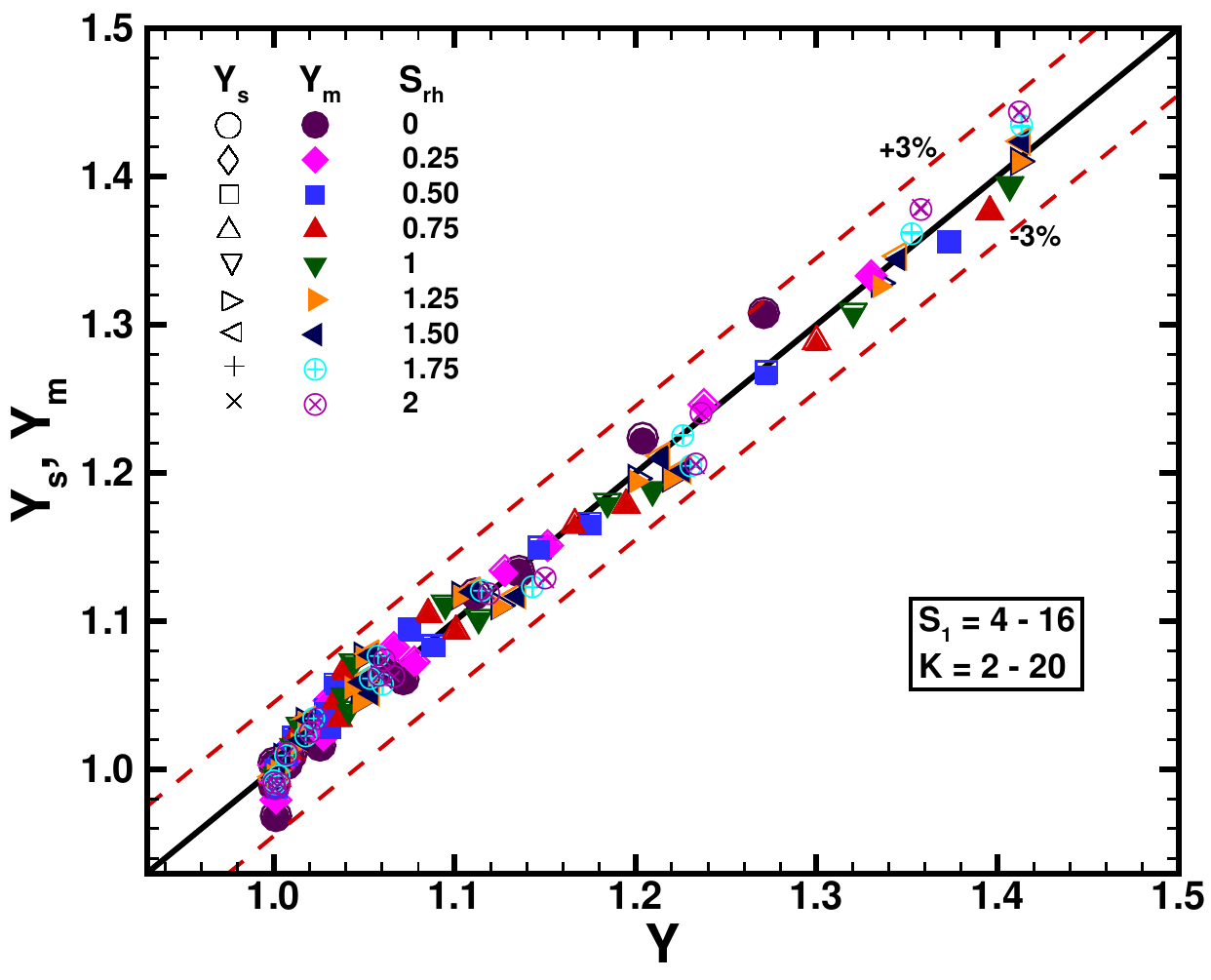}}
	\caption{Parity chart between numerically and mathematically (subscript m) obtained values of the (a) pressure drop, $\Delta P$ (\tab\ref{tab:1}) versus $\Delta P_\text{m}$ (\eqn\ref{eq:38}), and (b) electroviscous correction factor, $Y$ (\eqn\ref{eq:27} and \fig\ref{fig:13}) versus $Y_\text{s}$ (\eqn\ref{eq:y18}) and $Y_\text{m}$ (\eqn\ref{eq:40}) over the range of the flow governing parameters (\tab\ref{tab:1b}).}
	\label{fig:14}
\end{figure} 
%
\eqns(\ref{eq:33}) and (\ref{eq:38}) depict a generalized simpler pseudo-analytical model for the low Reynolds number ($Re$) flow of electrolyte liquid through a heterogeneously charged uniform slit microfluidic device. It is further extended to analytically calculate the electroviscous correction factor as follows.
\begin{gather}
	Y_{\text{m}}=\frac{\Delta P_{\text{m}}}{\Delta P_{0,\text{m}}} = \Gamma_\text{hr}
	\label{eq:40}
\end{gather}
%
\fig\ref{fig:14} presents the parity chart for pressure drop ($\Delta P$ vs $\Delta P_m$) and correction factor ($Y$ vs $Y_\text{m}$) obtained numerically and using simple pseudo-analytical model (\eqns\ref{eq:38} and \ref{eq:40}) for considered ranges of the flow conditions ($K$, $ {S_\text{1}}$, and $ {S_\text{rh}}$; \tab\ref{tab:1b}) in this study. A simple pseudo-analytical model approximates both pressure drop and electroviscous correction factor within $\pm3\%$ of present numerical values. The difference between a simple predictive model and present numerical results is reduced with decreasing surface charge density, surface charge-heterogeneity ratio, and EDL thickness. 

{In summary, pressure drop ($|\Delta P|$) increases with reducing $K$ and enhancing $S_\text{1}$. It is because $K$ is inversely proportional to the EDL thickness ($\lambda_\text{D}$); thus, decrement in $K$ augments the excess charge ($n^\ast$), which increases the electrical body force ($\textbf{F}_\text{e}$) and hence $|\Delta P|$ from momentum equation (\eqn\ref{eq:9}). The enhancement in $S_\text{1}$ increases the charge attractive force in the close vicinity of the walls, which imposes additional resistance on the flow and enhances $|\Delta P|$. Further, the increment in $S_\text{rh}$ increases electrostatic force near the walls of heterogeneous section due to enhancement in $S_\text{2}$ from \eqn(\ref{eq:9}) at fixed $S_\text{1}$. This enhanced electrostatic force decreases $n^\ast$ and increases $E_\text{x}$ in the device, intensifying the electrical force and further pressure drop ($|\Delta P|$). In addition, increment in $|\Delta P|$ increases the correction factor, $Y$ (maximally 41.35\%) from \eqn(\ref{eq:27}) with increasing $S_\text{rh}$. Thus, charge-heterogeneity enhances the electroviscous effects significantly in uniform geometries, which can be used to control and manipulate the practical microfluidic applications.}
%
\section{Concluding remarks}
\noindent 
This work has analyzed the electroviscous effects in the steady laminar pressure-driven flow of symmetric (1:1) electrolyte liquid through a symmetrically and heterogeneously charged uniform slit microfluidic device. The flow governing equations such as Poisson's, Nernst-Planck (NP), Navier-Stokes (NS) equations have been modeled using finite element method (FEM). The numerical results are presented in terms of the total electrical potential ($U$), excess charge ($n^\ast$), induced electric field strength ($E_\text{x}$), pressure ($P$), and electroviscous correction factor ($Y$) for the broad ranges of the electroviscous ($4\le S_\text{1}\le 16$, $0\le S_\text{rh}\le 2$, $2\le K\le 20$) and the non-electroviscous flow conditions ($K=\infty$, $S_\text{k} = 0$) at the low Reynolds number ($Re=0.01$).
Charge-heterogeneity ($S_\text{rh}$) complexly affects the hydrodynamic characteristics of the microfluidic device. The total electrical potential and pressure drop maximally change by 99.09\% (at $ {S_\text{1}}=4$ and $K=20$) and 12.77\% (at $ {S_\text{1}}=8$ and $K=2$), respectively when ${S_\text{rh}}$ varies from 0 to 2. The factor ($Y$) maximally increases by 12.77\% (at ${S_\text{1}}=8$ and $K=2$), 19.72\% (at ${S_\text{rh}}=0.5$ and $K=2$), and 40.98\% (at ${S_\text{rh}}=1.5$ and $ {S_\text{1}}=16$), respectively with the variation of ${S_\text{rh}}$ from 0 to 2, ${S_\text{1}}$ from 4 to 16, and $K$ from 20 to 2. Further, overall enhancement in $Y$ is recorded as 41.35\% at $K=2$, ${S_\text{1}}=16$, and ${S_\text{rh}}=1.5$, relative to non-EVE (nEVF, ${S_\text{k}}=0$ or $K=\infty$). 
Finally, a simple pseudo-analytical model is developed to calculate the pressure drop and electroviscous correction factor, which overpredicts the pressure drop (hence electroviscous correction factor) by $\pm3\%$ compared to numerical results. The difference between the predictive model and present numerical results values is reduced with increasing $K$ or EDL thinning and decreasing ${S_\text{1}}$ and ${S_\text{rh}}$. This generalized model and numerical correlations enable the present numerical results to be used to develop effective and reliable microfluidic devices for mixing, heat, and mass transfer processes for their practical applications.
%
\section*{Declaration of Competing Interest}
\noindent 
The authors declare that they have no known competing financial interests or personal relationships that could have appeared to influence the work reported in this paper.
%
\section*{Acknowledgements}
%
RPB would like to acknowledge Science and Engineering Research Board (SERB) - Department of Science and Technology (DST), Government of India (GoI) for the providence of the MATRICS grant (File no. MTR/2019/001598).  The authors acknowledge the infrastructural, computing resources, and software license support from the Indian Institute of Technology Roorkee. JD is thankful to the Department of Higher Education, Ministry of Education (MoE), Government of India (GoI) for the providence of research fellowship. 
%
\begin{spacing}{1.5}
\fontsize{10}{10pt}\selectfont
 \nomenclature[g0]{\textit{Greek letters}}{}
 \nomenclature[d0]{\textit{Dimensionless groups}}{}
 \nomenclature[s0]{\textit{Subscripts and Superscripts}}{}
 \nomenclature[z0]{\textit{Abbreviations}}{}
%
\nomenclature[zcfc]{CH}{charge-heterogeneity}
\nomenclature[zcfd]{CFD}{computational fluid dynamics}
\nomenclature[zedl]{EDL}{electrical double layer}
\nomenclature[zevf]{EVF}{electroviscous flow}
\nomenclature[zfem]{FEM}{finite element method}
\nomenclature[zpdes]{PDEs}{partial differential equations}
\nomenclature[zpdest]{PDF}{pressure-driven flow}
\nomenclature[zsaes]{SAEs}{simultaneous algebraic equations}
%
\nomenclature[aD]{$\mathcal{D}$}{diffusivity of the positive and negative ions, assumed equal ($\mathcal{D}_{+}=\mathcal{D}_{-}=\mathcal{D}$), m$^2$/s}
\nomenclature[aDj]{$\mathcal{D}_{j}$}{diffusivity of the ions of type j, m$^2$/s}
\nomenclature[ae]{$e$}{elementary charge of a proton ($=1.602176634\times 10^{-19}$), C or A.s}
\nomenclature[aE]{$E_{\text{x}}$}{induced electric field strength, V/m or --}
\nomenclature[afj]{$\mathbf{f_\text{j}}$}{flux density of the ions of type j (\eqn\ref{eq:7}), 1/(m$^2$.s)}
\nomenclature[aIc]{$I_{\text{c}}$}{conduction current density (\eqn\ref{eq:2}), A/m$^2$ or --}
\nomenclature[aId]{$I_{\text{d}}$}{diffusion current density (\eqn\ref{eq:2}), A/m$^2$ or --}
\nomenclature[aIs]{$I_{\text{s}}$}{streaming current density (\eqn\ref{eq:2}), A/m$^2$ or --}
\nomenclature[akB]{$k_{\text{B}}$}{Boltzmann constant ($=1.380649\times 10^{-23}$), J/K}
\nomenclature[aLc]{$L_{\text{h}}$}{length of heterogeneous section, m or --}
\nomenclature[aLd]{$L_{\text{d}}$}{length of downstream outlet section, m or --}
\nomenclature[aLu]{$L_{\text{u}}$}{length of upstream inlet section, m or --}
\nomenclature[an+]{$n_{+}$}{local number density of positive ions (\eqn\ref{eq:6}), 1/m$^3$ or --}
\nomenclature[an-]{$n_{-}$}{local number density of negative ions (\eqn\ref{eq:6}), 1/m$^3$ or --}
\nomenclature[an0]{$n_{0}$}{bulk density of the ions of type j, 1/m$^3$}
\nomenclature[anj]{$n_{j}$}{local number density of the ions of type j, 1/m$^3$}
\nomenclature[ans]{$n^*$}{excess charge ($=n_{+}-n_{-}$), 1/m$^3$ or --}
\nomenclature[aP]{$P$}{pressure, Pa or --}
\nomenclature[aT]{$T$}{temperature, K}
\nomenclature[aU]{$U$}{total electrical potential, V or --}
\nomenclature[aV]{$\mathbf{V}$}{velocity vector, m/s or --}
\nomenclature[aVa]{$\overline{V}$}{average velocity of the fluid at the inlet, m/s}
\nomenclature[aVx]{$V_x$}{x-component of the velocity, m/s or --}
\nomenclature[aVy]{$V_y$}{y-component of the velocity, m/s or --}
\nomenclature[aW]{$W$}{cross-sectional width of microchannel, m}
\nomenclature[ax]{$x$}{streamwise coordinate, --}
\nomenclature[ay]{$y$}{transverse coordinate, --}
\nomenclature[aY]{$Y$}{electroviscous correction factor (\eqns\ref{eq:27}, and \ref{eq:40}), --}
\nomenclature[azj]{$z_{j}$}{valency of the ions of type j, assumed equal ($z_{+}= -z_{-}=z$), --}
%
%
\nomenclature[gdP]{$\Delta P$}{pressure drop (\eqns\ref{eq:38}), --}
\nomenclature[geps0]{$\varepsilon_{\text{0}}$}{permittivity of free space (i.e. vaccum), F/m or C/(V.m)}
\nomenclature[gepsr]{$\varepsilon_{\text{r}}$}{dielectric constant (or absolute permittivity or relative permittivity) of the electrolyte liquid, --}
\nomenclature[glambdad]{$\lambda_{\text{D}}$}{Debye length $\left(=\sqrt{\frac{\varepsilon_{\text{0}}\varepsilon_{\text{r}} k_{\text{b}}T}{z^2e^2n_{\text{0}}}}\right)$, m}
\nomenclature[gmu]{$\mu$}{viscosity, Pa.s}
\nomenclature[gmueff]{$\mu_\text{eff}$}{effective or apparent viscosity, Pa.s}
\nomenclature[gpsi]{$\psi$}{EDL potential, V or --}
\nomenclature[grho]{$\rho$}{density of fluid, kg/m$^3$}
\nomenclature[grhoe]{$\rho_{\text{e}}$}{charge density of liquid, C/m$^3$}
\nomenclature[gsigmab]{$\sigma_\text{1}$}{upstream/downstream section surface charge density, C/m$^2$}
\nomenclature[gsigma]{$\sigma_\text{2}$}{heterogeneous section surface charge density, C/m$^2$}
%
%
\nomenclature[dbeta]{$\mathit{\beta}$}{liquid parameter (\eqn\ref{eq:14}), --}
\nomenclature[dK]{$\mathit{K}$}{inverse Debye length (\eqn\ref{eq:14}), --}
\nomenclature[dPe]{$Pe$}{Peclet number ($={Re}~\mathit{Sc}$) (\eqn\ref{eq:14}), --}
\nomenclature[dRe]{$Re$}{Reynolds number (\eqn\ref{eq:14}), --}
\nomenclature[dSa]{$\mathit{S_\text{2}}$}{heterogeneous section surface charge density (\eqn\ref{eq:3}), --}
\nomenclature[dSc]{$\mathit{Sc}$}{Schmidt number (\eqn\ref{eq:14}), --}
\nomenclature[dSb]{$\mathit{S_\text{rh}}$}{surface charge-heterogeneity ratio (\eqn\ref{eq:4}), --}
\nomenclature[dS]{$\mathit{S_\text{1}}$}{upstream/downstream section surface charge density (\eqn\ref{eq:3}), --}
%
\nomenclature[sz]{$0$}{without electroviscous effects}
\nomenclature[sd]{$d$}{downstream}
\nomenclature[se]{$e$}{extra or excess
\nomenclature[sf]{$h$}{heterogeneous}}
\nomenclature[sm]{$m$}{mathematical}
\nomenclature[ss]{$s$}{statistical}
\nomenclature[su]{$u$}{upstream}
%

\printnomenclature
\end{spacing}
%
%
%
%
\bibliography{references}
%
%
%
%
%
%
%
%
%
%
%
%
\end{document}